\documentclass[useAMS,usenatbib,usegraphicx]{mn2e}

\title[Super-Orbital Variability in Hard X-rays]{Super-Orbital Variability in Hard X-rays}

\author[S. A. Farrell et al.]{S. A. Farrell$^{1}$\thanks{E-mail: sean.farrell@cesr.fr}, D.
Barret$^{1}$ and G. K. Skinner $^{2,3}$\\
$^{1}$Centre d'Etude Spatiale des Rayonnements, CNRS/UPS, 9 Avenue du Colonel Roche, 31028 Toulouse Cedex 4, France\\
$^{2}$CRESST and Astroparticle Physics Laboratory, Code 661,
NASA/Goddard Space Flight Centre, Greenbelt, MD 20771, USA\\
$^{3}$Department of Astronomy, University of Maryland, College
Park, MD 20742, USA\\}

\begin{document}

\date{Released 2008 Xxxxx XX}

\pagerange{\pageref{firstpage}--\pageref{lastpage}} \pubyear{2008}

\maketitle

\label{firstpage}

\begin{abstract}
We present the results of a study with the \emph{Swift} Burst
Alert Telescope in the 14 -- 195 keV range of the long-term
variability of 5 low mass X-ray binaries with reported or
suspected super-orbital periods --- 4U 1636-536, 4U 1820-303, 4U
1916-053, Cyg X-2 and Sco X-1. No significant persistent periodic
modulation was detected around the previously reported periods in
the 4U 1916-053, Cyg X-2 or Sco X-1 light curves. The $\sim$170 d
period of 4U 1820-303 was detected up to 24 keV, consistent with
variable accretion due to the previously proposed triple system
model. The $\sim$46 d period in 4U 1636-536 was detected up to 100
keV, with the modulation in the low and high energy bands found to
be phase shifted by $\sim$180$^\circ$ with respect to each other.
This phase shift, when taken together with the near-coincident
onset of the $\sim$46 d modulation and the low/hard X-ray state,
leads us to speculate that the modulation could herald transient
jet formation.

\end{abstract}

\begin{keywords}
accretion, accretion discs  -  stars: neutron  -  X-rays: binary
\end{keywords}

\section{Introduction}

Long-term ``super-orbital'' periodic variability has been observed
in soft X-rays from over 30 X-ray binaries with the All Sky
Monitor (ASM) on the \emph{Rossi X-ray Timing Explorer}
(\emph{RXTE}) satellite \citep[see for example][]{so07}. Also
known as `long' or `third' periods, these modulations are defined
simply as any periodic variability greater than the orbital
period. This very general definition of super-orbital variability
is consistent with a number of physical mechanisms for the
variation. Unlike binary orbital or neutron star spin periods, the
mechanisms behind super-orbital variability are not well
understood. The precession of a warped accretion disc modulating
the X-ray flux from the compact object is currently the favoured
model \citep[see for example][]{wij99,cl03a}. This warp can be
caused by the gravitational field of the donor star \citep[see for
example][]{la97}, or by radiation pressure exerted by the central
X-ray emitting regions \citep[see for example][]{wij99}.
\citet{og01} showed that while radiation driven warping gives a
coherent picture of super-orbital periods, not all systems agree
with its predictions \citep[e.g. 2S 0114+650;][]{fa07}.

Other models proposed to explain super-orbital variability include
variations in the accretion rate \citep[caused by stellar
pulsations of the donor star, irregular Roche lobe overflow or
asymmetric stellar wind capture;][]{we92,ps06}, precession of the
neutron star spin-axis in an X-ray pulsar \citep*{tr86,sta00},
variations in the location of an accreting hot-spot \citep*{ru92}
and the presence of a third body in the system \citep*[i.e. a
triple system;][]{pa00,ch01}.

While most super-orbital periods are at best quasi-periodic, four
X-ray binaries have been shown to exhibit stable super-orbital
modulation over long timescales -- Her X-1, LMC X-4, SS 433 and 2S
0114+650 \citep{pa02,so06}. Other systems such as SMC X-1 have
super-orbital periods which have been observed to vary smoothly
within a well-defined range \citep{cl03a}, while other
super-orbital periods appear to evolve stochastically. It is thus
clear that there is more than one mechanism behind the
super-orbital modulations.

A number of X-ray binaries were regularly monitored in hard X-rays
by the \emph{Compton Gamma Ray Observatory} (\emph{CGRO}) Burst
and Transient Source Experiment (BATSE) for the first $\sim$1,600
days of the \emph{RXTE} mission \citep{har04}. Since the
decommissioning of \emph{CGRO} in June 2000, the lack of high
energy response of the \emph{RXTE} ASM has precluded long-term
studies of super-orbital variability simultaneously at low and
high energies. The launch of \emph{Swift} in 2004 saw the arrival
of a wide field monitoring instrument capable of complementing the
\emph{RXTE} ASM in hard X-rays -- the Burst Alert Telescope
\citep[BAT;][]{bar05}. In this paper we present the results of a
search for the previously reported super-orbital modulation of 5
low mass X-ray binaries (LMXBs) in the BAT 14 -- 195 keV light
curves.

\section{Sources}

\subsection{4U 1636-536}

4U 1636-536 was discovered following a series of type I X-ray
bursts observed with the \emph{SAS 3} satellite \citep*{ho77} and
later classified as an atoll source \citep{ha89}. The orbital
period of 3.8 hr was derived from optical photometric measurements
by \citet*{pe81}. A distance of 4.3 $\pm$ 0.6 kpc was derived by
\citet{jon04} using type I photospheric radius expansion bursts.
However, more recent analyses by \citet{gal06} derived a distance
of 6.0 $\pm$ 0.5 kpc for a canonical neutron star (i.e. M$_{NS}$ =
1.4 M$_\odot$, R$_{NS}$ = 10 km).

\citet{ca06} performed phase-resolved VLT spectroscopy of the
irradiated donor in 4U 1636-536, using Doppler imaging of the NIII
$\lambda$4640 Bowen transition lines to determine a new set of
spectroscopic ephemerides. A reanalysis of burst oscillations
using these ephemerides allowed them to place the first dynamical
constraints on the masses of the binary companions, constraining
the mass function (f(M) = 0.76 $\pm$ 0.47 M$_\odot$), mass-ratio
(q $\sim$ 0.21 -- 0.34) and system inclination angle (i $\sim$ 36
-- 60$^\circ$).

\citet{sh05} reported the onset of a large amplitude,
statistically significant periodic modulation at $\sim$46 d in the
\emph{RXTE} ASM data, interpreting this periodicity as due to an
accretion rate variability related to the X-ray irradiation of the
disc. These same authors noted a gradual decline in the X-ray flux
of this source coincident with the appearance of the $\sim$46 d
variability, arguing that 4U 1636-53 was transiting from activity
to quiescence due to a reduction in the mass accretion rate.
\citet{fi06}, however, observed a hard tail dominating the
spectrum above 30 keV using data from the \emph{INTEGRAL}
satellite, speculating that this high energy emission could be
linked to the onset of relativistic outflows (i.e. jets).

The hard state of atoll sources has been previously considered to
be associated with the presence of a steady jet and radio emission
\citep[see for example][]{fe06,mi06}. When compared with black
hole systems containing jets, \citet{mi06} found that the neutron
star systems were less radio loud for a given X-ray luminosity,
but that the jet power in both classes of object scales linearly
with accretion rate. Neutron star LMXBs in the low/hard state are
thus expected to contain weaker radio jets than their black hole
counterparts.

A number of radio observations of 4U 1636-536 were performed
before the onset of the 46 d modulation, the decreased X-ray flux,
and the hardening of the X-ray spectrum, showing an absence of
significant radio emission (see \S~6.2.4).  No radio observations
targeting this source after the onset of the modulation or the
transition to the low/hard state have been reported in the
literature. As such, the presence of jets in this system is yet to
be confirmed.

\subsection{4U 1820-303}
4U 1820-303 was discovered near the centre of the globular cluster
NGC 6624 in 1974 by the \emph{Uhuru} satellite \citep*{gi74}. The
subsequent discovery of intense bursts associated with
thermonuclear flashes from this source confirmed it as a low mass
X-ray binary containing a neutron star, making this the first
identified source of type I X-ray bursts \citep*{gr76}. A distance
of 7.6 $\pm$ 0.4 kpc has been estimated using the bursts as a
standard candle \citep*{ku03}.

The system consists of a white dwarf and a neutron star orbiting
their common centre of mass with an orbital period of 11.4 min
\citep*{ra87,st87}. \citet*{ha89} later classified 4U 1820-303 as
an atoll source, with \citet*{zh98} observing the presence of kHz
quasi-periodic oscillations (QPOs). \citet{bl00b} observed that
the QPO frequency was strongly correlated with the position on the
colour-colour diagram. This result (in combination with the
observed saturation of the QPO frequency at high accretion rates)
lead them to conclude that they were observing the last stable
orbit in the accretion disc.

\citet{pr84b} noted a periodic $\sim$170 d variability in the
X-ray flux, over which timescale the source moves from the
soft/banana state to the hard/island state. This quasi-periodic
$\sim$170 d variability was later proposed to be a result of
variable mass accretion due to tidally induced modulation of the
orbital eccentricity by a third massive body \citep{ch01}.
Recently, \citet{ta07} reported the detection of a hard X-ray tail
above 50 keV from this source in the island state, concluding that
this emission is likely due either to emission of non-thermal
electrons or thermal emission from plasmas with a relatively high
temperature.

\citet*{zd07a} performed an analysis of the super-orbital
modulation using \emph{RXTE} ASM data. These authors undertook a
study of hierarchical triple system models yielding the required
quasi-periodic eccentricity oscillations, finding that the
resulting theoretical light curves matched well with the observed
ones. \citet{zd07b} reported a dependence on the X-ray spectral
state of both the depth and phasing of the orbital X-ray
modulation. This led them to favour a model in which the observed
X-ray modulation is caused by scattering in hot gas around an
accretion-stream bulge at the disc edge; as the mass accretion
rate changes with the $\sim$170 d cycle, so the size and location
of this bulge varies. The average flux calculated over the course
of the super-orbital variability was found to be compatible with
the model of accretion due to the angular momentum loss via
emission of gravitational radiation.

\subsection{4U 1916-053}

4U 1916-053 was discovered in 1974 by the \emph{Uhuru} satellite
\citep{gi74} and is the most compact X-ray binary to exhibit
periodic intensity dips. Type I X-ray bursts each lasting $\sim$10
s have been observed from this source with a recurrence timescale
of several hours \citep{be77,le77}. A distance of 8.8 $\pm$ 1.3
kpc has been estimated by \citet{jon04} from photospheric radius
expansion during the bursts. 4U 1916-053 is classified as an atoll
source and displays a similar path in the X-ray color-color
diagram to 4U 1820-303, although with significantly lower
luminosity \citep{bl00a}.

The X-ray dips have been ascribed to obscuration of the X-ray
emitting region by a bulge in the outer accretion disk where the
accretion stream meets the disc \citep{wh82}. The dip recurrence
period of $\sim$3000 s is the shortest among the known dipping
sources \citep*{sw84}. The optical counterpart was discovered as a
V = 21 mag optical star modulating at a period of 3027.5 s by
\citet{gr88}. This optical period is $\sim$1\% longer than the
X-ray dip period, raising the question as to which of the values
indicated the orbital period.

A 3.9 d beat period between the optical and X-ray periods was
reported by \citet{gr92}, manifesting as a modulation of the X-ray
dip shape \citep*{ch01b}. Timing analyses of \emph{RXTE} PCA data
and quasi-simultaneous optical data performed by \citet{ch01b}
lead them to conclude that the X-ray dip period was indeed the
binary orbital period, with the slightly longer optical period
produced through beating between the orbital period and the 3.9 d
period, which they ascribed to the precession period of the
accretion disc. They also set an upper limit of 2.06 $\times$
10$^{-11}$ s s$^{-1}$ on the $|\dot{P}|$, arguing that the
inferred stability and the partial clustering of X-ray bursts
preceding the X-ray dips continues to suggest that the system may
be a hierarchical triple.

A regular long-term periodicity of 199 d (and a possible period of
40.6 d) was reported in the X-ray flux using data from the
\emph{Vela 5B} satellite by \citep{pr84a}, which the authors
ascribed to disc precession. Later analysis of the same data by
\citet{sm92} found that the 199 d modulation was not very strong,
with a confidence level of just 80\%. \citet{ho01} analysed
$\sim$3.7 yr of \emph{RXTE} ASM data, finding no significant peaks
above the red or white noise levels at any period. \citet{si05}
investigated the long-term X-ray activity of 4U 1916-053 using 7
yr of \emph{RXTE} ASM data, finding that a cycle of about 230 --
250 d occurred only in a limited time interval, materialising as
variations of the recurrence time and amplitude of outbursts.
\citet{si05} also concluded that the movement of 4U 1916-053 along
the atoll path in the X-ray color-color diagram during the
outbursts implied that the observed long-term variability was
caused by variable accretion onto the neutron star, and not by
disk precession (which would imply occultation and/or absorption
effects). A more recent analysis of 8.5 yr of \emph{RXTE} ASM data
by \citet{we06} failed to detect any long-term periodic
variability.

\subsection{Cyg X-2}

Cyg X-2 is one of the six Galactic LMXBs that are classified as
Z-sources \citep{ha89}, and was first discovered by \citet*{by66}
with a sounding rocket experiment. Type I X-ray bursts were
reported by \citet{sm98}, confirming a neutron star as the compact
object. Distances of 7.2 $\pm$ 1.1 kpc and 13 $\pm$ 2 kpc have
been estimated from optical observations \citep{or99} and from
photospheric radius expansion in the bursts \citep{jon04}
respectively. A binary orbital period of 9.8444 d was deduced from
the optical behaviour of the donor star \citep*{ca98}. A high
energy tail has been observed in the X-ray spectrum during a
number of observations \citep*{di02,pi02,la06}, although hard
X-ray emission does not appear to be a constant feature of this
source \citep{la06}.

A 77 d periodicity in data from the \emph{Vela 5B} satellite was
reported by \citet{sm92}. \citet*{wi96} reported the detection of
a 78 d period in \emph{RXTE} ASM data from the first 160 d of the
mission, and showed that their result was supported by archival
data from the \emph{Ariel 5} and \emph{Vela 5B} missions. However,
analyses of archived data from the all sky monitors on the
\emph{RXTE}, \emph{Ginga}, \emph{Ariel 5}, and \emph{Vela 5B}
satellites and the scanning modulation collimator of \emph{HEAO 1}
by \citet{pa00} found no evidence for a stable modulation in the
X-ray emission. Strong peaks at 40.4 d and 68.8 d in the
\emph{RXTE} ASM data, and at 53.7 d and 61.3 d in the \emph{Ginga}
ASM data were observed by these authors, leading them to conclude
that the super-orbital variability in this system (if real) is
aperiodic. Dynamic analyses performed on 7 yr of \emph{RXTE} ASM
data by \citet{cl03b} revealed complex quasi-periodic behaviour,
with a highly significant periodicity at $\sim$59 d in a 300 d
section of the light curve, with significant recurring features at
$\sim$40 d appearing sporadically.

\subsection{Sco X-1}

The Z-source Sco X-1 was the first X-ray binary discovered and is
the brightest persistent extra-solar X-ray source \citep{gi62}.
Although the system properties are consistent with it containing a
neutron star with a weak magnetic field, no type I bursts have
been observed so far \citep{mi03}. Radio emission from the
vicinity of Sco X-1 was first reported by \citet{an68}, with later
high-resolution radio interferometry unveiling compact jets
through the formation of twin radio lobes with variable morphology
\citep*{fo01}. A distance of 2.8 $\pm$ 0.3 kpc was determined from
trigonometric parallax \citep*{br99}. The orbital period of
$\sim$18.9 hr was originally discovered in 85 yr of archived
photometric optical data by \citet*{go75}, and more recently
observed in X-rays by \citet*{va03}.

A high energy tail above 30 keV in the X-ray spectrum was observed
in \emph{CGRO} OSSE data by \citet{st00} and later verified in
\emph{RXTE} and \emph{INTEGRAL} data by \citet{da01} and
\citet{di06} respectively. A periodic modulation at 37 d was
reported by \citet{pe96} in 9 months of \emph{RXTE} ASM data,
although this modulation has yet to be confirmed.

\section{Data Reduction}

While awaiting gamma-ray bursts the \emph{Swift} BAT instrument
accumulates data in survey mode. A given source is within the
field of view of the BAT  $\sim$10\% of the time, with a typical
integration time per pointing of around 1000 s. Standard software
was used to extract and clean 'detector plane images' in four
energy bands for each 'observation' (typically a few thousand
seconds, often comprising several shorter segments) and to
calculate the arrays of weights corresponding to the visibility of
each source within the field of view by each detector pixel during
that observation.  A program developed for the purpose by the
authors was then used to fit simultaneously the intensities of the
sources and a series of background model components (equivalent
functionality to that provided by this program is now available in
the latest public software, though the public software was not
available at the time the analyses were performed). Corrections
were applied for earth occultation and the known energy-dependent
variation of the instrument sensitivity with off-axis angle.

\section{Data Analysis}

\begin{table*}
 \centering
 \begin{minipage}{130mm}
\caption{Swift BAT and $\emph{RXTE}$ ASM light curve properties
for each of the four sources analysed.}\label{tab1}
\begin{tabular}{ccccccccccc}
\hline  Source & \multicolumn{2}{c}{\emph{RXTE} ASM} & \multicolumn{2}{c}{BAT A}& \multicolumn{2}{c}{BAT B}& \multicolumn{2}{c}{BAT C}& \multicolumn{2}{c}{BAT D}\\
& N\footnote{number of points in the light curve} & f$_{samp}$\footnote{average light curve sampling frequency in units of d$^{-1}$}& N & f$_{samp}$& N & f$_{samp}$& N & f$_{samp}$& N & f$_{samp}$\\
\hline
4U 1636-536& 481& 0.77&483 &0.78 &485 &0.78 &485 &0.78 &494& 0.80\\
4U 1820-303& 452&0.72 &419 &0.75 &405 &0.72 &426 &0.76 &429 &0.76\\
4U 1916-053& 485& 0.78&379 &0.69 &384 &0.70 &386 &0.70 & 390&0.71\\
Cyg X-2&570& 0.91&504 & 0.81&493 &0.79 &507 & 0.81&507 &0.81\\
Sco X-1& 485& 0.77&483 & 0.77&470 & 0.75&466 &0.74 &465 &0.74\\
\hline
\end{tabular}
\end{minipage}
\end{table*}

The BAT 14 -- 24 keV (A), 24 -- 50 keV (B), 50 -- 100 keV (C) and
100 -- 195 keV (D) light curves were used for these analyses,
rebinned at 1 d. The average sampling rate for the reduced light
curves prior to rebinning varied from $\sim$0.8 -- 12.5 d$^{-1}$.
For comparison, the \emph{RXTE} ASM 1.5 -- 12 keV 1 d average
light curves for the same time span were analysed to cover the low
energy range of the spectrum. The 1 d binning was chosen for
convenience as we were interested only in the long-term
variability of the target sources. Table \ref{tab1} lists the
number of data points and average sampling frequencies for the BAT
and \emph{RXTE} ASM light curves for each source.

All data points were weighted using the modified weighting scheme
developed by \citet*{co07}. In this method, individual data points
are given weights based on both the non-uniform measurement
uncertainties and the intrinsic source variability. The resulting
weighted count rate $R'_{i}$ is given by:
\begin{eqnarray}
R'_{i} =  \frac{R_i}{[(f \sigma_i)^2 + V_s]} \label{eq1}
\end{eqnarray}
where $R_{i}$ is the raw un-weighted rate, $f$ is a correction
factor to account for underestimation of the error on each point,
and $\sigma_i$ is the uncertainty for the point. $V_s$ is the
estimated variance due to source variability, given by:
\begin{eqnarray}
V_s = \frac{\sum^{N}_{i=1}(\overline{R} - R_i)^2}{N - 1} -
\frac{\sum^{N}_{i=1}(f \sigma_i)^2}{N} \label{eq2}
\end{eqnarray}
where $\overline{R}$ is the mean count rate and $N$ is the number
of data points in the light curve.

The $f$ factor was derived for each of the energy bands in the BAT
and \emph{RXTE} ASM light curves by fitting a constant to the
light curves of a source expected to be stable. The $f$ factor was
determined by taking the square root of the reduced $\chi^2$ value
obtained from fitting data from the Crab (R. H. D. Corbet 2007,
private communication). For cases in which the $V_s$ parameter was
negative (applicable to sources which are very faint in the given
energy band), $V_s$ was set to zero and the method reverts to the
``simple'' weighting technique proposed by \citet{sc89}.

Power spectra were generated for each of the weighted light curves
using the fast periodogram \textsc{fasper} subroutine of the
Lomb-Scargle periodogram \citep{lo75,sc82,pr89}. The over sampling
and hi-frequency factors -- which set the period range and
resolution \citep{pr89} -- were both set at 2 for each analysis
run, allowing us to search for periodic variability in the
$\sim$1.3 -- 1000 d range. The Lomb-Scargle technique allows us to
sample below the average Nyquist period (with reduced sensitivity)
due to the unevenly sampled nature of the data, without creating
problems due to aliasing as would be seen in the analysis of an
evenly sampled data set using standard fast Fourier techniques.
The frequency range and binning were chosen to give good coverage
of the range covered by known super-orbital periods while at the
same time providing sufficient resolution in the power spectra.
The 99$\%$ white noise significance levels were estimated using
Monte Carlo simulations \citep*[see for example][]{ko98}. The
99$\%$ red noise significance levels were estimated using the
$\textsc{redfit}$
subroutine\footnote{ftp://ftp.ncdc.noaa.gov/pub/data/paleo/softlib/redfit/redfit\_preprint.pdf},
which fits a first-order autoregressive process to the time series
in order to estimate the red-noise spectrum \citep{sc02}.

\section{Results}

\subsection{4U 1636-536}

Figure \ref{4U16lc} shows the \emph{RXTE} ASM and \emph{Swift} BAT
light curves for 4U 1636-536 while Figure \ref{4U16} shows the
power spectra. The previously reported 46 d periodicity is clearly
detected in the \emph{RXTE} ASM 1.5 -- 12 keV data, and is also
significantly present in the BAT 14 -- 24 keV, 24 -- 50 keV and 50
-- 100 keV light curves. No significant modulation above 100 keV
is detected in the 1.3 -- 1000 d range. In order to investigate
the likely origin of the 46 d modulation, the raw (i.e.
un-weighted) \emph{RXTE} ASM, BAT A, BAT B and BAT C light curves
were each folded over the 46 d period, setting phase zero
arbitrarily to the start of the BAT light curves at MJD 53360. The
profile of the modulation was observed to shift in phase between
the 1.5 -- 12 keV and 14 -- 24 keV energy bands, so that the
profiles above 14 keV are almost 180$^\circ$ out of phase with the
lower energy modulation.

\begin{figure}
\includegraphics[width=\columnwidth]{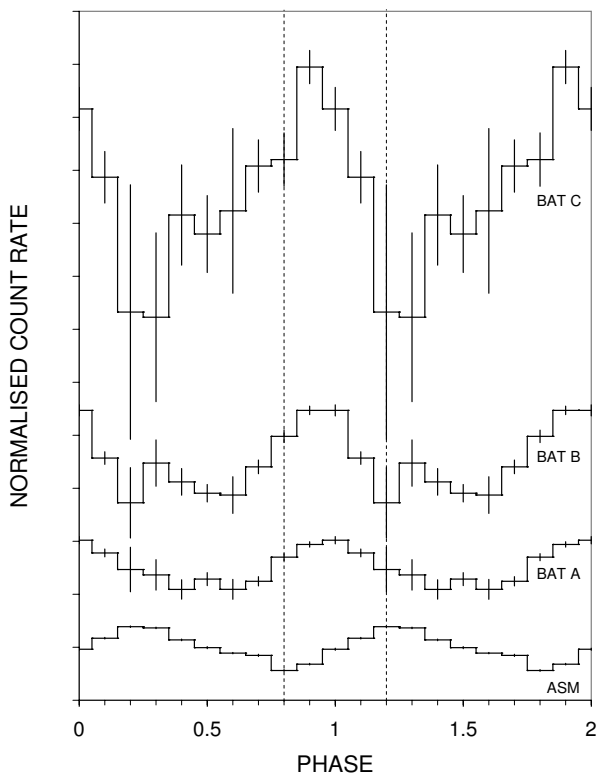}
\caption{The light curves of 4U 1636-536 folded over the 46 d
period. From bottom: \emph{RXTE} ASM 1.5 -- 12 keV profile, BAT A
14 -- 24 keV profile, BAT B 24 -- 50 keV profile, and BAT C 50 --
100 keV profile. The vertical dashed lines indicate the phases
corresponding to the minimum and maximum of the \emph{RXTE} ASM
profile. The count rates are normalised to the mean of each light
curve and the scale for each profile is the same.}\label{16fold}
\end{figure}

It can be clearly seen in Figure \ref{4U16lc} that the period of
the modulation is not stable, with the time between successive
peaks varying over time. In order to test how the drift in the
period affects the width of the peak in the power spectra, we
simulated light curves with the same sampling as the \emph{RXTE}
ASM, BAT A, BAT B and BAT C light curves with a stable sinusoidal
modulation with a period of 46 d. Power spectra were then
generated and compared with those produced using the real data
shown in Figure \ref{4U16}. In all cases the peaks in the real
power spectra were marginally wider than the simulated spectra.
While this discrepancy could possibly be linked to the change in
the period, the difference could also be due to the effects of
noise or statistical scatter.

Analysis of the BAT light curves prior to rebinning into 1 d
average data sets found an additional peak at $\sim$77 above the
99$\%$ white noise significance levels. \citet*{fa05} have shown
that by imposing a known periodic sampling rate, any spurious
peaks in the power spectrum produced by aliasing or spectral
leakage should be shifted to other frequencies. By simulating
light curves with periodic modulations at 46 and 77 d with the
same sampling as the BAT data we found that rebinning the light
curves should increase the significance of a real peak. The
significance of the 77 d peak was dramatically reduced in power
spectra of the rebinned BAT light curves (see Figure \ref{4U16})
and is well below the dominant 99$\%$ white noise significance
levels. We thus identify the 77 d peak as spurious.

\subsection{4U 1820-303}

\begin{figure}
\includegraphics[width=\columnwidth]{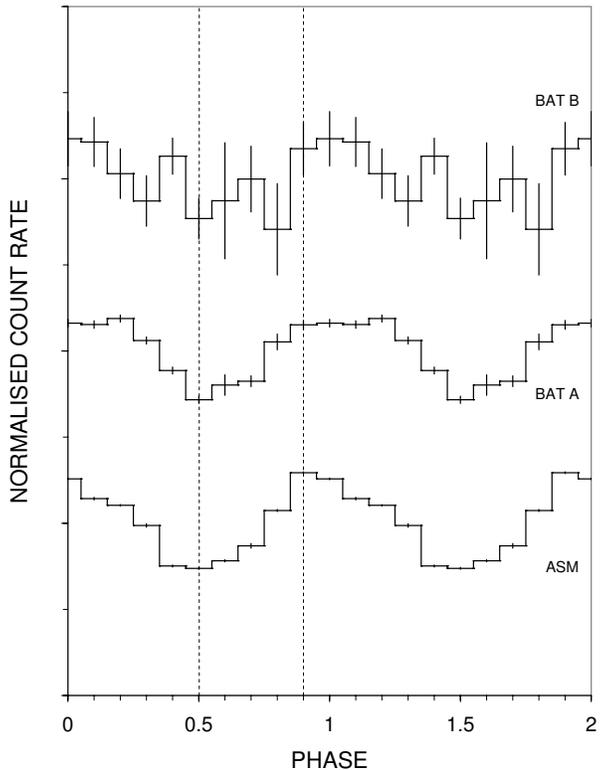}
\caption{The light curves of 4U 1820-303 folded over the 170 d
period. From bottom: \emph{RXTE} ASM 1.5 -- 12 keV profile, BAT A
14 -- 24 keV profile and BAT B 24 -- 50 keV profile. The vertical
dashed lines indicate the phases corresponding to the minimum and
maximum of the \emph{RXTE} ASM profile. The count rates are
normalised to the mean of each light curve and the scale for each
profile is the same.}\label{18fold}
\end{figure}

Figure \ref{4U18lc} shows the \emph{RXTE} ASM and \emph{Swift} BAT
light curves for 4U 1820-303 while Figure \ref{4U18} shows the
power spectra. The previously reported $\sim$170 d periodicity is
clearly detected in the \emph{RXTE} ASM 1.5 -- 12 keV data as a
single broad peak, and is also significantly present in the BAT 14
-- 24 keV light curves. No peaks around 46 d are significantly
present in the 24 -- 195 keV power spectra. The 1.5 -- 12 keV and
14 -- 24 keV light curves were thus epoch folded over the 170 d
period, setting phase zero arbitrarily to the start of the ASM
light curve at MJD 53353. The count rates were normalised to the
mean rate (Figure \ref{18fold}). Unlike 4U 1636-536, no phase
shift was observed between the low and high energy profiles.

The period of the super-orbital modulation of 4U 1820-303 has been
known to vary over time \citep{zd07a}. As such, to test the effect
of a varying period on the width of the peak in the power spectra,
we simulated light curves with a stable sinusoidal modulation at
170 d with the same sampling as the \emph{RXTE} ASM and BAT A
light curves. We then generated power spectra for the simulated
light curves and compared them with the power spectra shown in
Figure \ref{4U18}. As with 4U 1636-536, the peaks in the real
power spectra were marginally wider than those generated using the
simulated data. While this discrepancy could again be linked to a
changing period, there is no more than a suggestion of possible
slight broadening of the peaks.

The simulations clearly show that the width of the peak is a
function of both the period of the modulation and the number of
cycles in the data set. Therefore, in order to compare the width
of the super-orbital peak in the 4U 1820-303 power spectra with
that of 4U 1636-536 this relationship needs to be taken into
account. We thus took the full width of the peaks at 50$\%$ of the
maximum power, normalised it by the period of the modulation and
then multiplied by the number of cycles in the light curves.
Comparing the normalised peak widths shown in Table \ref{widths},
we see that the values for the real peaks are comparable for both
4U 1636-536 and 4U 1820-303, and are consistently marginally
greater than the width of the simulated peaks.

\begin{table}
\begin{center}
\caption[]{Comparison between the normalised widths of the
super-orbital peaks in real and simulated power spectra for 4U
1636-536 and 4U 1820-303.}\label{widths}
\begin{tabular}{lcccc}
\hline Data Set& \multicolumn{2}{c}{4U 1636-536}& \multicolumn{2}{c}{4U 1820-303}\\
&W$_{real}$&W$_{sim}$&W$_{real}$&W$_{sim}$\\
\hline
\emph{RXTE} ASM & 0.94& 0.82& 1.11&0.90\\
BAT A & 1.07& 0.88& 0.94& 0.86\\
BAT B & 0.97& 0.88& -- &-- \\
BAT C & 1.00&0.88 &-- &-- \\
 \hline
\end{tabular}
\end{center}
\end{table}

In addition to the $\sim$170 d modulation, a peak at around
$\sim$100 d marginally exceeds the 99$\%$ white noise significance
level in the 24 -- 50 keV power spectrum. However, the lack of a
significant peak at this period in the other BAT power spectra (in
particular between 14 -- 24 keV) leads us to conclude that this
peak is spurious. No other significant modulation above 50 keV is
detected in the 1.3 -- 1000 d range.

\subsection{4U 1916-053}
Figure \ref{4U19lc} shows the \emph{RXTE} ASM and \emph{Swift} BAT
light curves for 4U 1916-053 while Figure \ref{4U19} shows the
power spectra. No significant modulation in any of the light
curves is apparent in the 1.3 -- 1000 d range, indicating that
either the 199 d and 40.6 d periodicities detected by
\citet{pr84a} were spurious, the mechanism for these modulations
was not active during the time of our observations, or that the
signal was not sufficiently stable and/or persistent over the
entire observing window for it to be detected in a single power
spectrum.

\begin{figure}
\includegraphics[width=\columnwidth]{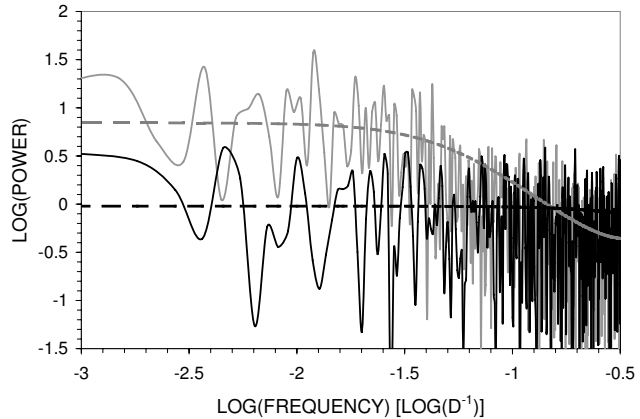}
\caption{Log-log power spectra of the 1.5 -- 12 keV \emph{RXTE}
ASM light curves of 4U 1916-053 covering the \citet{ho01} time
frame (grey solid lines) and the time frame covered by the BAT
data (black solid lines). The red noise models for each power
spectrum are plotted as dashed lines for
comparison.}\label{1916rn}
\end{figure}

The power spectrum of the \emph{RXTE} ASM data for 4U 1916-053
appears to be dominated by white noise. This result is in contrast
to the results of \citet{ho01}, who found that red noise dominated
the power spectrum of the first four years of \emph{RXTE} ASM
data. Analysis of the \citet{ho01} data using our analysis methods
confirm that red noise was indeed dominant in the \emph{RXTE} ASM
data, but has since dropped in significance. Figure \ref{1916rn}
presents a comparison between the power spectra generated for the
\emph{RXTE} ASM data covering the \citet{ho01} epoch (MJD 50150 --
51500) and the time span used for our analyses (MJD 53360 --
53900), clearly showing the drop in red noise.

\subsection{Cyg X-2}

Figure \ref{cygx2lc} shows the \emph{RXTE} ASM and \emph{Swift}
BAT light curves for Cyg X-2 while Figure \ref{CygX2} shows the
power spectra. A strong, broad peak at $\sim$190 d is present in
the \emph{RXTE} ASM 1.5 -- 12 keV and BAT 14 -- 24 keV power
spectra. These peaks are located at approximately 0.5 yr, a common
location for spurious peaks \citep[see for see for
example][]{fa05}. No other significant peaks are present in any of
the energy bands. A peak at $\sim$40 d is present in the power
spectrum from the \emph{RXTE} ASM data, similar to the 40.4 d
signal detected by \citet{pa00}. Although the power of this peak
exceeds the 99$\%$ white noise significance level, it is well
below the red noise significance level.

\subsection{Sco X-1}

Figure \ref{scox1lc} shows the \emph{RXTE} ASM and \emph{Swift}
BAT light curves for Sco X-1 while Figure \ref{ScoX1} shows the
power spectra. No periodic variability greater than the dominant
99$\%$ white noise significance level is present in the
\emph{RXTE} ASM or BAT power spectra in the 1.3 -- 1000 d range.

\section{Discussion}

The typical shape of the X-ray spectrum for LMXBs consists of a
thermal Comptonisation component (sometimes empirically modelled
as a power law with exponential cut-off) curtailed at low energies
by photoelectric absorption, often with the addition of a
low-energy black body component arising either from thermal
emission from the inner regions of the disc or very close to the
surface of the neutron star \citep[see for example][for a
review]{bl00b,bar01}. In the low/hard state an additional power
law component is sometimes required in order to model the hard
X-ray tail \citep[see for example][]{ta07}, with the hard X-ray
emission attributed variously to Comptonisation by
high-temperature thermal or non-thermal electrons in the accretion
disc corona or in compact jets.

\subsection{The $\sim$170 d Period in 4U 1820-303}
\begin{table*}
 \centering
 \begin{minipage}{140mm}
\caption{Derived flux upper limits for 4U 1820-303 in the island
state in the BAT B, C and D energy bands using the model
parameters in \citet{bl00b} and \citet{ta07}.}\label{tab2}
\begin{tabular}{lccccc}
\hline  BAT Energy Band & \multicolumn{5}{c}{Flux Upper Limits ($\times$ 10$^{-11}$ ergs cm$^{-2}$ s$^{-1}$)}\\
& CPL + BB\footnote{cut-off power law plus black body model \citep{bl00b}} & CompTT + BB\footnote{thermal Comptonisation plus black body model \citep{bl00b}} & CPL + DBB\footnote{cut-off power law plus disk black body model \citep{bl00b}} & CompTT\footnote{thermal Comptonisation model \citep{ta07}} & CompTT + PL\footnote{thermal Comptonisation plus power law model \citep{ta07}}\\
\hline
B (24 -- 50 keV) &5.5   & 5.5 & 5.2 & 4.8 &5.5 \\
C (50 -- 100 keV) &1.5  & 1.5 &1.6  &1.6  &1.7 \\
D (100 -- 195 keV) &7.4  &7.0  &7.6  &7.9  &8.6 \\
\hline
\end{tabular}
\end{minipage}
\end{table*}

The $\sim$170 d period in 4U 1820-303 has been attributed to
variable accretion linked to the presence of a third body in this
system, whereby the eccentricity of the inner orbit is modulated
by tidal effects produced by the outer body \citep{ch01,zd07a}.
Transitions between the high/soft and low/hard states over this
timescale are linked to variability in the accretion rate
\citep{bl00b} and the movement in and out of the inner region of
the accretion disc \citep[see for example][]{mal07}, thus
modulating the thermal Comptonisation region of the spectrum. This
thermal Comptonisation component has been seen to dominate the
spectrum up to $\sim$50 keV, above which 4U 1820-303 has only been
detected on one occasion during the low/hard (island) state, during
which a high energy non-thermal tail was found to dominate
\citep{ta07}.

The $\sim$170 d period has been seen to correspond to normal atoll
source motion in the colour-colour diagram, with the minimum of
the modulation coinciding with the transition to the island state
\citep{bl00b}. If the appearance of the high energy non-thermal
tail is regular and tied to the island state (and thus the minimum
of the super-orbital modulation), a hardening of the spectrum
would be anticipated every $\sim$170 d. We would therefore expect
to detect the modulation above 50 keV, anti-correlated with the
low energy modulation. If the hard tail in 4U 1820-303
preferentially appears on a regular basis during the transition to
the island state, we would also expect the low and high energy
modulations to be out of phase. However, as mentioned in $\S$ 5.2,
no such phase shift is apparent in the folded profiles (Figure
\ref{18fold}).

The super-orbital period should be present at least up to the 50
keV threshold, as a varying accretion rate would be expected to
modulate the entire thermal Compton region of the spectrum. To
test whether 4U 1820-303 would be detected by the BAT above 24
keV, we simulated spectra (1000 s integration times) using the BAT
response files for the various island state spectral models
applied to \emph{RXTE} and \emph{INTEGRAL} data by \citet{bl00b}
and \citet{ta07}, respectively.

Flux upper limits during the minimum of the $\sim$170 d modulation
(i.e. the island state) in the three BAT energy bands above 24 keV
were calculated for each of the \citet{bl00b} and \citet{ta07}
island state spectral models. Table \ref{tab2} lists the derived
flux upper limits in the BAT B, C and D energy bands for each of
the \citet{bl00b} and \citet{ta07} models. The projected flux in
each band for each spectral model was also calculated, with the
normalisation parameters set so that the 14 -- 24 keV model count
rate equalled the measured rate in this band (the only BAT band in
which we can be certain 4U 1820-303 was detected).

In all cases except the \citet{ta07} simple Comptonisation and the
\citet{bl00b} cut-off power law + disc black body models, the
expected flux was far in excess of the derived upper limits in the
24 -- 50 keV and 50 -- 100 keV bands. Thus, within the
uncertainties of this method, the non-detection of the $\sim$170 d
modulation above 24 keV is most consistent with the softest island
state spectral models in \citet{ta07} and \citet{bl00b}.

\subsection{The $\sim$46 d Period in 4U 1636-536}

\subsubsection{Formation of Compact Jets?}

\begin{figure}
\includegraphics[width=\columnwidth]{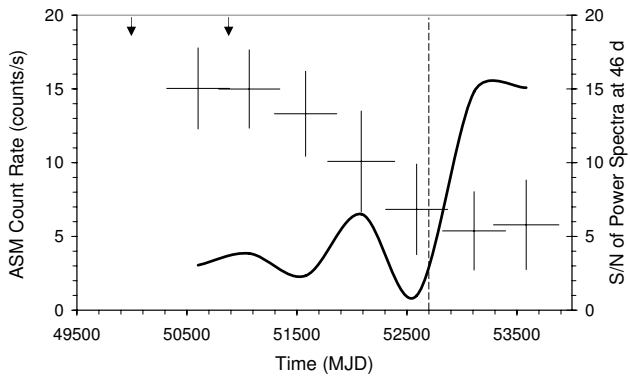}
\caption{Average \emph{RXTE} ASM 1.5 -- 12.0 keV count rates
(crosses) from 4U 1636-536 compared with the signal-to-noise of
the 46 d peak in the power spectra (solid line). The arrows
indicate the times of two previous radio observations made with
the ATCA, while the dashed line indicates the timing of the
\emph{INTEGRAL} observations which first detected the high energy
tail in the X-ray spectrum.}\label{flux2}
\end{figure}

\citet{fi06} linked the recent reduction in X-ray flux and the
appearance of a high energy tail in the spectrum of 4U 1636-536 to
the onset of jet formation. Interestingly, the 46 d modulation
first appears in the \emph{RXTE} ASM data around the same time as
this transition to the low/hard state. Figure \ref{flux2} shows
the coarse binned \emph{RXTE} ASM light curve up to MJD 53900 for
4U 1636-536 with the signal-to-noise (S/N) of the 46 d peak (as
taken from overlapping power spectra generated from 600 d light
curve sections with a 100 d overlap). The significance of the
modulation can clearly be seen to increase markedly with the
decrease in flux (the significance of the 46 d periodicity did not
exceed the 99$\%$ significance levels until a S/N of $\sim$7 was
reached). The appearance of a high energy power law tail in the
spectrum in conjunction with the first detection of the 46 d
period and the reduction in flux implies a link between the three
phenomena. If the assertion made by \citet{fi06} is correct (that
the transition to the low/hard island state is linked to jet
formation), the 46 d modulation could be tied to transient jet
formation, or possibly to the precession of jets (linked in turn
to the precession of a warped disc) in a similar fashion as seen
in SS 433 \citep{ma84}.

As mentioned in $\S$ 5.1, the $\sim$46 d modulation is clearly
seen in each of the folded profiles shown in Figure \ref{16fold},
with the phase of the modulation shifting with increasing energy
so that the BAT C (50 -- 100 keV) profile is almost 180$^\circ$
out of phase with the \emph{RXTE} ASM (1.5 -- 12 keV) profile. A
similar phase shift was observed between the \emph{RXTE} ASM and
\emph{INTEGRAL} light curves by \citet{sh05}, which was attributed
to the high and low energy modulations originating at different
locations. In comparison, the profiles of the light curves of 4U
1820-303 when folded over the 170 d period show no such phase
shift (Figure \ref{18fold}).

\subsubsection{Interpreting the Phase Shift}

The detection of the 46 d modulation in both the high and low
energy realms of the spectrum indicate that both the thermal and
non-thermal components are varying over this timescale. The phase
shift of the modulation is inconsistent with the standard warped
precessing disc model for super-orbital variability, as periodic
obscuration or reflection would be expected to cause the high and
low energy modulations to vary in phase. We interpret this phase
shift as an indication that the 46 d modulation arises from two
separate (yet closely linked) phenomena. If the low energy
modulation seen in the \emph{RXTE} ASM light curves is indeed due
to variable obscuration of the central emission region (either the
inner regions of the disc or the neutron star itself), and the
high energy modulation seen in the BAT data is a result of a
varying viewing geometry of the beamed emission from the jet, it
is possible to see two similar modulations in different energy
bands that are out of phase. In this scenario, the jet column will
be pointing towards us at roughly the same time as the bulge in
the disc is obscuring the central emission region. Thus we would
see an increase in the high energy flux at the same time as a
decrease in the low energy flux (as observed in the \emph{RXTE}
ASM and BAT data), with the high and low energy flux varying
smoothly as the disc and jet precess together.

Many systems which show super-orbital modulation ascribed to disc
precession do not show any evidence for jet formation, and in some
cases jet formation is inhibited due to the presence of a highly
magnetised neutron star (e.g. Her X-1). However, if a warped
precessing disc is present, any jets which form would expect to be
coupled to the warped disc and should themselves precess. The jet
would be aligned with the normal axis of the inner disc, so the
scenario in which the bulge is phased to obscure the inner disc at
the precise time as the jet points towards us is unlikely.

\subsubsection{Transient Jet Formation?}

A somewhat simpler explanation would be that the 46 d modulation
is linked to periodic transitions between the high/soft and
low/hard states, leading to transient jet formation. In this
alternate scenario, the inner region of the disc moves in as the
accretion rate intensifies, resulting in an increase in the flux
of seed photons and therefore an enhancement in the thermal
Comptonisation (i.e. soft) component of the spectrum. In this
high/soft state the conditions are not believed to be sufficient
for jet formation: the disc is thought to be geometrically thin
yet optically thick, with a relatively small corona as most of the
accretion power is believed to be dissipated in the disc itself
\citep[see for example][]{mal07}. The hard region of the spectrum
during this state is characterised by a weak, steep power law,
generally interpreted as inverse Compton up-scattering of soft
photons by non-thermal electrons in the corona \citep[see for
example][]{bar01}. As the accretion rate decreases, the inner
regions of the disc move further out and the disc cools, becoming
geometrically thick yet optically thin. In this state most of the
accretion power is believed to be transported into the corona,
which in turn is thought to power the compact jet. The soft
thermal component from the disc decreases as the flux of seed
photons lessens, and the strong hard non-thermal component appears
in conjunction with the jets. Thus an anti-correlation between the
hard and soft X-ray flux would result, with both fluxes modulated
over the $\sim$46 d period. In this scenario the 46 d period would
represent the characteristic timescale between disruption and
recovery of the disc.

\subsubsection{Searching for Radio Emission}

Recently, \citet{mig07} showed there to be a direct link between
radio jet emission and the X-ray state in the neutron star X-ray
binary GX 17+2. They also showed that the presence of a hard X-ray
tail was coupled to the low/hard X-ray state, implying a link
between the presence of radio jets and the appearance of the hard
tail in the spectrum. Thus, if transient jet formation linked to
spectral state transitions is the cause of the 46 d period, we
would expect radio emission from this source to be modulated in
phase with the hard X-ray modulation, as the jets form and radio
emission ensues \citep[see for example][]{cor03}.

A review of the Australian Telescope Compact Array (ATCA) online
archive (ATOA) shows that two previous radio observations of 4U
1636-536 were performed on 1995 September 2 and 1998 April 4, with
upper limits of 0.12 mJy and 0.15 mJy derived at 4.8 GHz and 8.6
GHz respectively \citep{ber00}. Both observations were performed
during the high/soft state, when compact jets and resulting radio
emission would not be expected. More recent observations were
performed with the ATCA on 2007 June 29, 30 and July 1 as part of
a multi-wavelength collaboration to study LMXBs of all kinds at
various
epochs\footnote{http://www.cfa.harvard.edu/twiki/bin/view/GreatObs/JetsDisks}.
While this data is still proprietary, the principal investigator
of these observations was kind enough to communicate the
preliminary results; a 3$\sigma$ upper limit of $\sim$0.6 mJy was
derived at 8.6 GHz (J. Dickey 2007, private communication).

The 2 -- 10 keV X-ray flux in the low/hard state measured with
\emph{INTEGRAL} was $\sim$1 $\times$ 10$^{-9}$ erg cm$^{-2}$
s$^{-1}$ (M. Fiocchi 2007, private communication), giving a
luminosity of $\sim$2 $\times$ 10$^{36}$ erg s$^{-1}$ for a
distance of 4.3 kpc \citep{jon04} and $\sim$4 $\times$ 10$^{36}$
erg s$^{-1}$ for a distance of 6 kpc \citep{gal06}. According to
the X-ray/radio luminosity correlation for neutron star LMXBs
derived by Migliari $\&$ Fender (2006), 4U 1636-536 should thus
have a radio luminosity (from measurements around 8.5 GHz) of
somewhere between (3 -- 20) $\times$ 10$^{28}$ erg s$^{-1}$. The
8.5 GHz flux density would thus be $\sim$1 mJy, barely sufficient
for a detection to be made in the recent ATCA observations.

\begin{figure}
\includegraphics[width=\columnwidth]{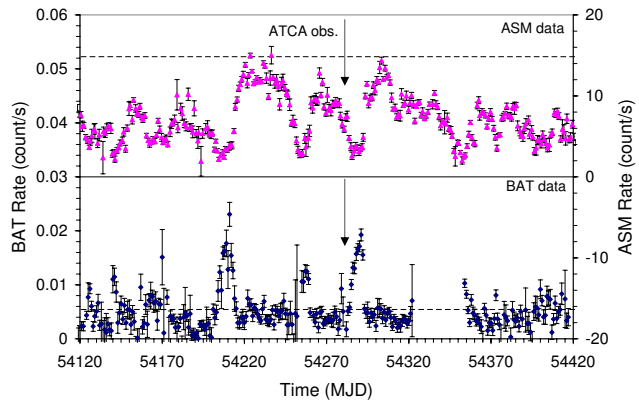}
\caption{\emph{RXTE} ASM (top) and BAT 14 -- 50 keV (bottom) light
curves covering MJD 54220 -- 54420, showing the timing of the
recent ATCA observations of 4U 1636-536 (arrows), which took place
just prior to the transition to the low/hard island state. The
dashed horizontal lines indicate the average BAT count rate and
the average high/soft state count rate in the \emph{RXTE} ASM. The
latter line shows that the source re-enters the high/soft state at
the peak of the $\sim$46 d modulation in the soft X-ray
band.}\label{1636_lcs}
\end{figure}

\begin{figure}
\includegraphics[width=\columnwidth]{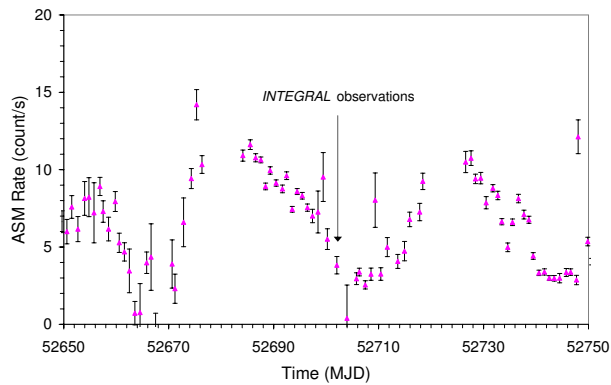}
\caption{The \emph{RXTE} ASM light curve of 4U 1636-536 covering
MJD 52650 -- 52750, showing the timing of the \emph{INTEGRAL}
observations (arrow) which detected the hard X-ray tail. These
observations took place during the drop in low energy flux, which
presumably coincided with a rise in the high energy flux. As the
observations took place after the decommissioning of \emph{CGRO}
and prior to the launch of \emph{Swift}, long-term monitoring data
in hard X-rays is not available during this epoch.}\label{int_tim}
\end{figure}

Comparing the timing of these recent observations with the
\emph{RXTE} ASM 1.5 -- 12 keV and the BAT 14 -- 50 keV light
curves (Figure \ref{1636_lcs}) shows that they occurred just after
the maximum of the modulation in the soft X-ray band, and just
before the increase in hard X-ray flux. These observations thus
appear to have been performed before the transition to the
low/hard state and the predicted phase of jet formation. Further
support for this theory is obtained when the timing of the
\emph{INTEGRAL} observations
--- during which the hard tail was detected by \citet{fi06} --- is
compared with the soft X-ray \emph{RXTE} ASM light curve (Figure
\ref{int_tim}). It can clearly be seen that the detection of the
hard tail occurred during the minimum of the soft X-ray
modulation, at the same time as we would have expected to see a
peak in the BAT hard X-ray light curve had monitoring data been
available during this epoch.

\subsection{Non-detections}

No evidence for the previously reported super-orbital periods in
4U 1916-053, Cyg X-2 or Sco X-1 was found in the \emph{RXTE} ASM
or any of the BAT light curves. The power spectra were instead
dominated by low-frequency noise, with no peaks that would
indicate the presence of real long-term periodic modulation. The
modulations for 4U 1916-053 and Cyg X-2 were originally reported
using \emph{Vela 5B} data, with later analyses by \citet{we06} of
8.5 yr of \emph{RXTE} ASM data finding no significant stable
modulation from either source. The detection of super-orbital
modulation in Sco X-1 was reported using only the first 9 months
of \emph{RXTE} ASM data. It is thus apparent that either the
reported detections were spurious, the mechanisms behind the
super-orbital variability in each of these three systems has since
ceased, or the modulation does not persist over a sufficient
fraction of the observing window to be detectable by our
techniques. Unfortunately, the BAT light curves do not cover a
sufficiently long epoch for us to subdivide the light curves, thus
precluding dynamic analyses to search for signals that are
potentially transient or variable in frequency.

\section{Conclusions}

We have analysed the \emph{RXTE} ASM and \emph{Swift} BAT light
curves in the 1.5 -- 195 keV range for 4U 1636-536, 4U 1820-303,
4U 1916-053, Cyg X-2 and Sco X-1. We found no evidence for
periodic variability at the previously reported super-orbital
periods from 4U 1916-053, Cyg X-2 or Sco X-1 in either the
\emph{RXTE} ASM or BAT data. In contrast, the $\sim$170 d period
in 4U 1820-303 is clearly detected up to 24 keV, consistent with
the modulation arising from variable accretion due to tidal
effects from the third body in the triple system. The 46 d period
in 4U 1636-536 was significantly detected in the \emph{RXTE} ASM
data and up to 100 keV in the BAT data, with the profile of the
high energy modulation found to be $\sim$180$^\circ$ out of phase
with that seen in the 1.5 -- 12 keV band. This behaviour is
inconsistent with the modulation arising solely from variable
photoelectric absorption due to the precession of a warp in the
accretion disc, leading us to interpret it as evidence that the
high and low energy modulations arise from two separate yet
closely linked phenomena. The appearance of the 46 d period in
conjunction with the transition to the low/hard state is
intriguing, leading us to speculate that this modulation could be
tied to the formation of transient compact jets. In this scenario,
the low energy modulation would result from a thermal Compton
component of the spectrum which varies with the accretion rate.
The high energy modulation would thus be tied to the appearance of
the non-thermal hard tail arising from the formation of compact
jets during episodes of decreased accretion. Further monitoring
observations at radio wavelengths are being sought in order to
confirm this conclusion.

\section*{Acknowledgments}

This work made use of quick-look results provided by the
ASM/\emph{RXTE} team. The authors thank members of the
\emph{Swift} BAT team for helpful discussions. We thank Steven
Tingay for his valuable discussions and John Dickey for kindly
communicating the preliminary results of the ATCA observations of
4U 1636-536. We thank the anonymous referee for their comments
which improved this paper.

\appendix
\section{Light Curves}

\begin{figure*}
\includegraphics[width=180mm]{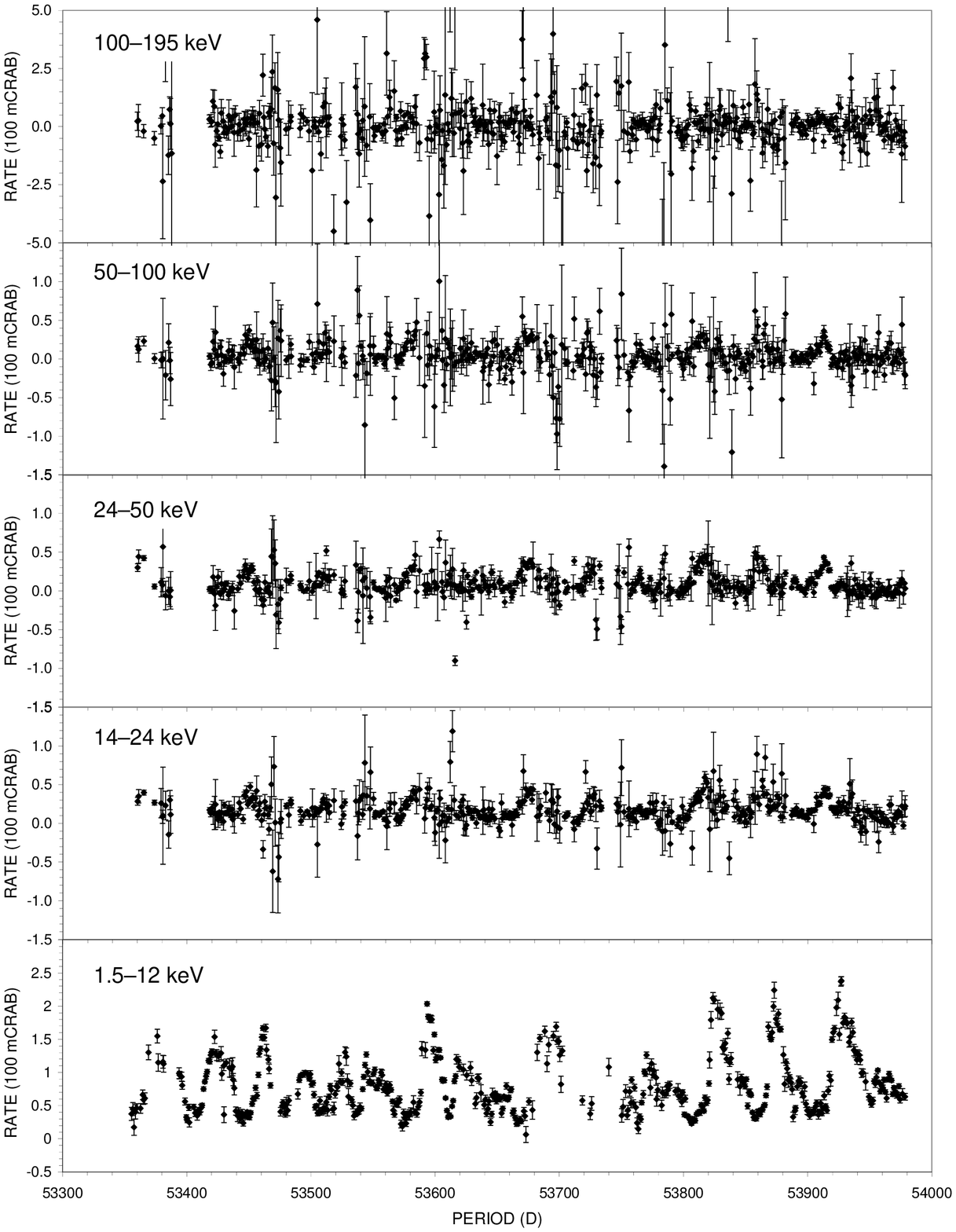}
\caption{4U 1636-536: the 1.5 -- 12 keV \emph{RXTE} ASM light
curve (bottom panel) and the 14 -- 24 keV, 24 -- 50 keV, 50 -- 100
keV and 100 -- 195 keV \emph{Swift} BAT light curves (top panels).
}\label{4U16lc}
\end{figure*}

\begin{figure*}
\includegraphics[width=180mm]{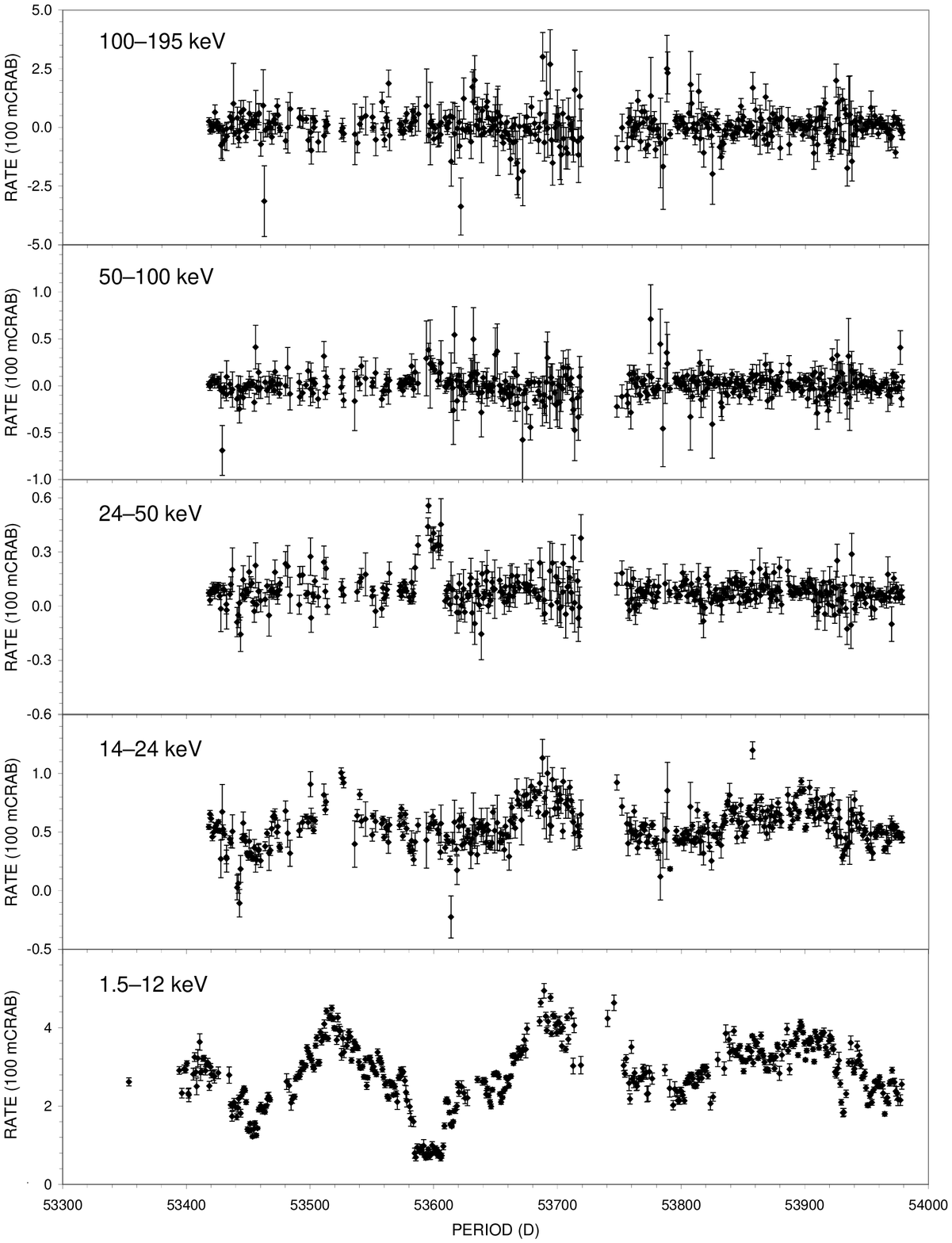}
\caption{4U 1820-303: the 1.5 -- 12 keV \emph{RXTE} ASM light
curve (bottom panel) and the 14 -- 24 keV, 24 -- 50 keV, 50 -- 100
keV and 100 -- 195 keV \emph{Swift} BAT light curves (top
panels).}\label{4U18lc}
\end{figure*}

\begin{figure*}
\includegraphics[width=180mm]{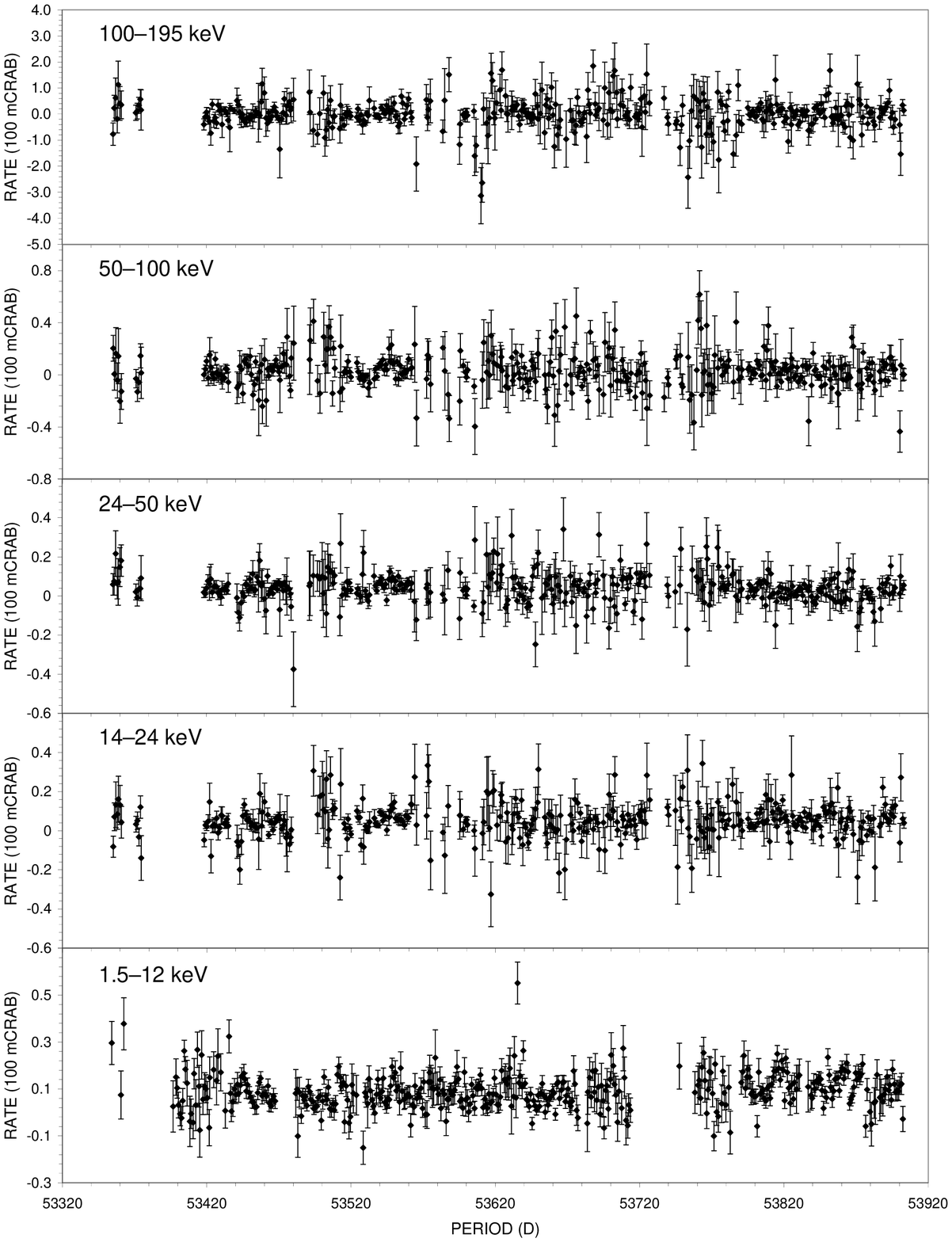}
\caption{4U 1916-053: the 1.5 -- 12 keV \emph{RXTE} ASM light
curve (bottom panel) and the 14 -- 24 keV, 24 -- 50 keV, 50 -- 100
keV and 100 -- 195 keV \emph{Swift} BAT light curves (top panels).
}\label{4U19lc}
\end{figure*}

\begin{figure*}
\includegraphics[width=180mm]{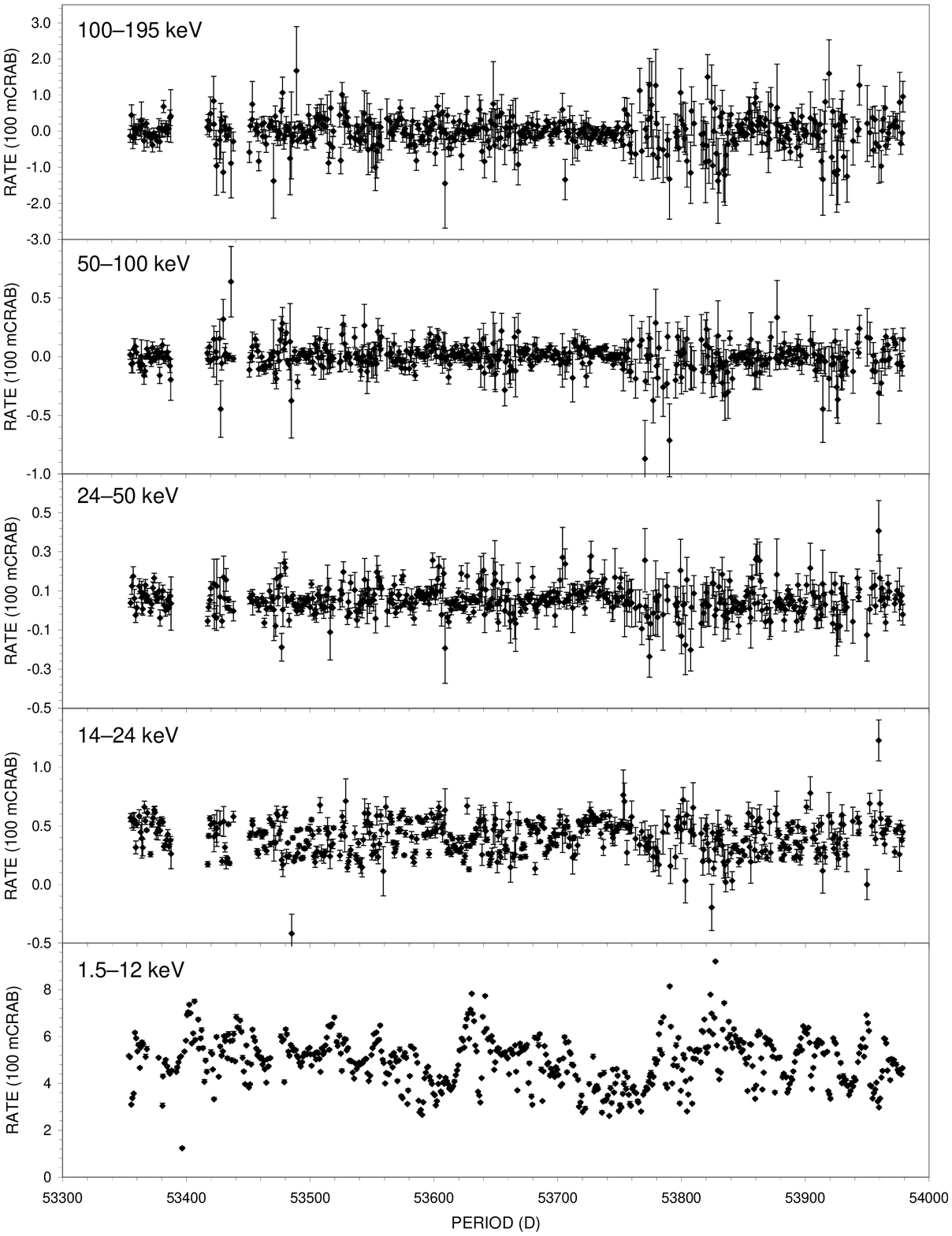}
\caption{Cyg X-2: the 1.5 -- 12 keV \emph{RXTE} ASM light curve
(bottom panel) and the 14 -- 24 keV, 24 -- 50 keV, 50 -- 100 keV
and 100 -- 195 keV \emph{Swift} BAT light curves (top
panels).}\label{cygx2lc}
\end{figure*}

\begin{figure*}
\includegraphics[width=180mm]{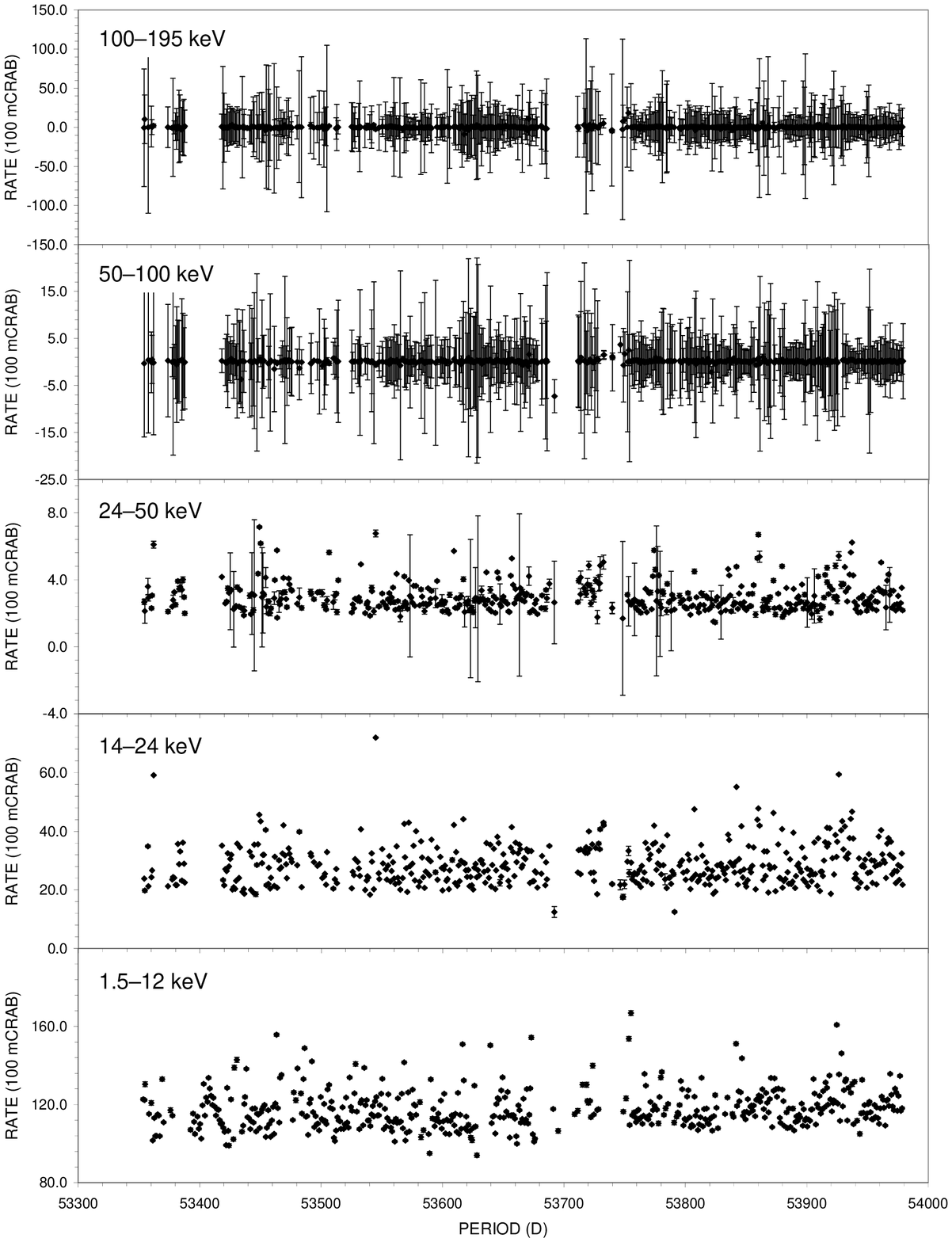}
\caption{Sco X-1: the 1.5 -- 12 keV \emph{RXTE} ASM light curve
(bottom panel) and the 14 -- 24 keV, 24 -- 50 keV, 50 -- 100 keV
and 100 -- 195 keV \emph{Swift} BAT light curves (top panels).
}\label{scox1lc}
\end{figure*}

\section{Power Spectra}
\clearpage

\begin{figure*}
\includegraphics[width=140mm]{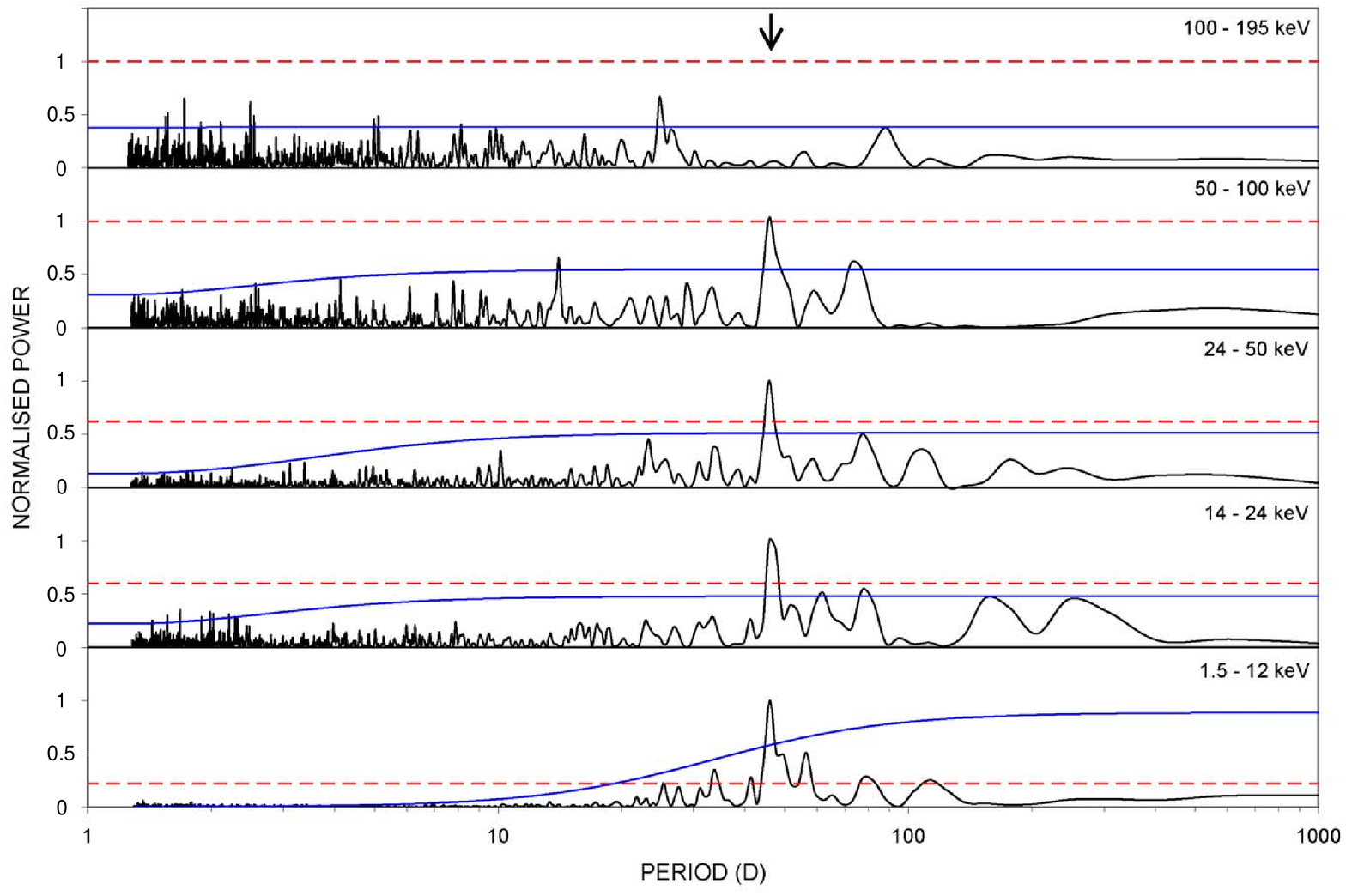}
\caption{4U 1636-536: normalised power spectra of the 1.5 -- 12
keV \emph{RXTE} ASM light curve (bottom panel) and the 14 -- 24
keV, 24 -- 50 keV, 50 -- 100 keV and 100 -- 195 keV \emph{Swift}
BAT light curves (top panels). The red horizontal dashed lines and
blue solid lines indicate the 99$\%$ white and red noise
significance levels respectively. The arrow indicates the location
of the previously reported super-orbital period of 46 d. The power
spectra were each normalised to the maximum power present, the
99$\%$ white or red noise levels, whichever was
greatest.}\label{4U16}
\end{figure*}

\begin{figure*}
\includegraphics[width=140mm]{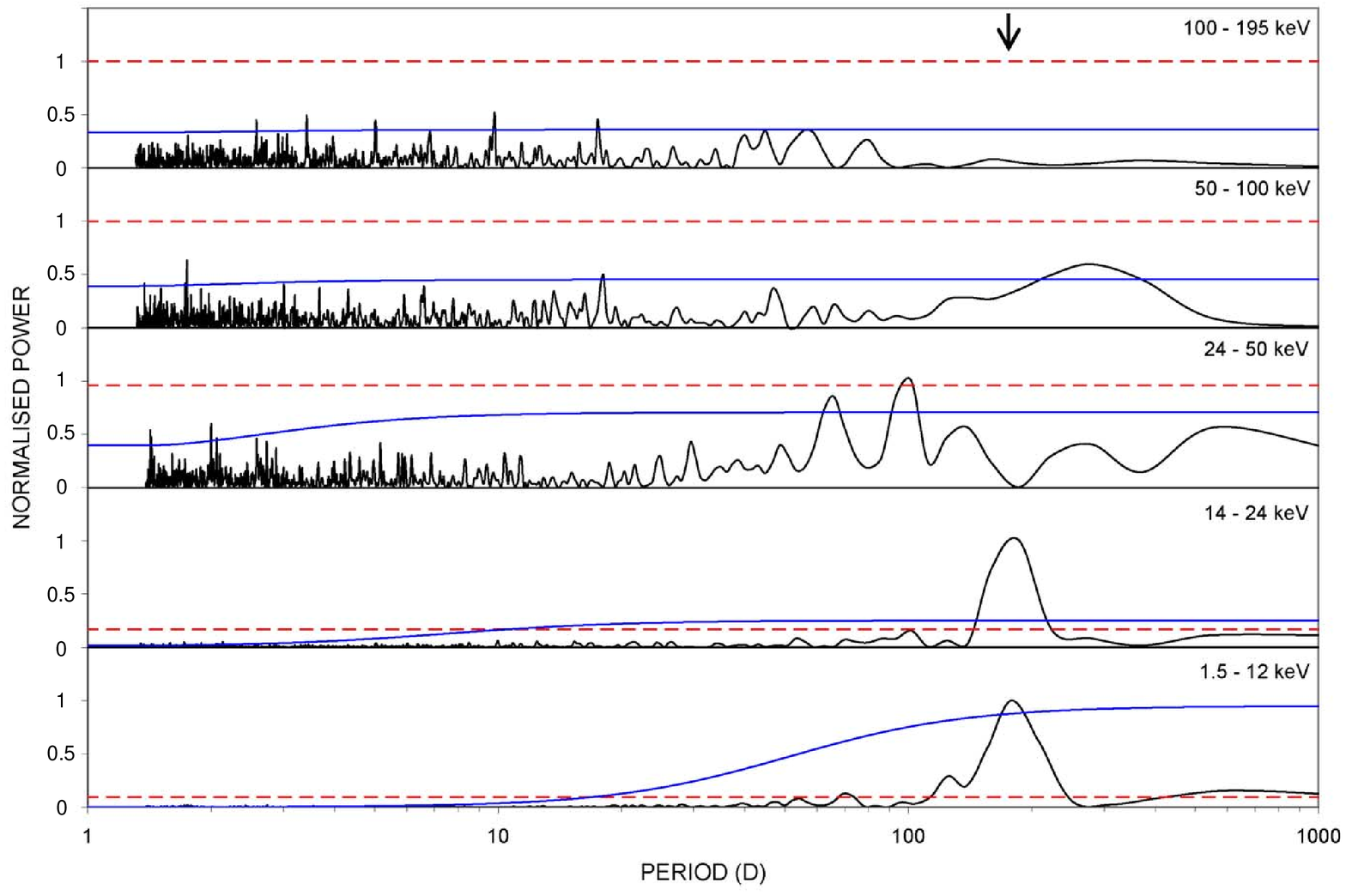}
\caption{4U 1820-303: normalised power spectra of the 1.5 -- 12
keV \emph{RXTE} ASM light curve (bottom panel) and the 14 -- 24
keV, 24 -- 50 keV, 50 -- 100 keV and 100 -- 195 keV \emph{Swift}
BAT light curves (top panels). The arrow indicates the location of
the previously reported super-orbital period of $\sim$170
d.}\label{4U18}
\end{figure*}

\begin{figure*}
\includegraphics[width=140mm]{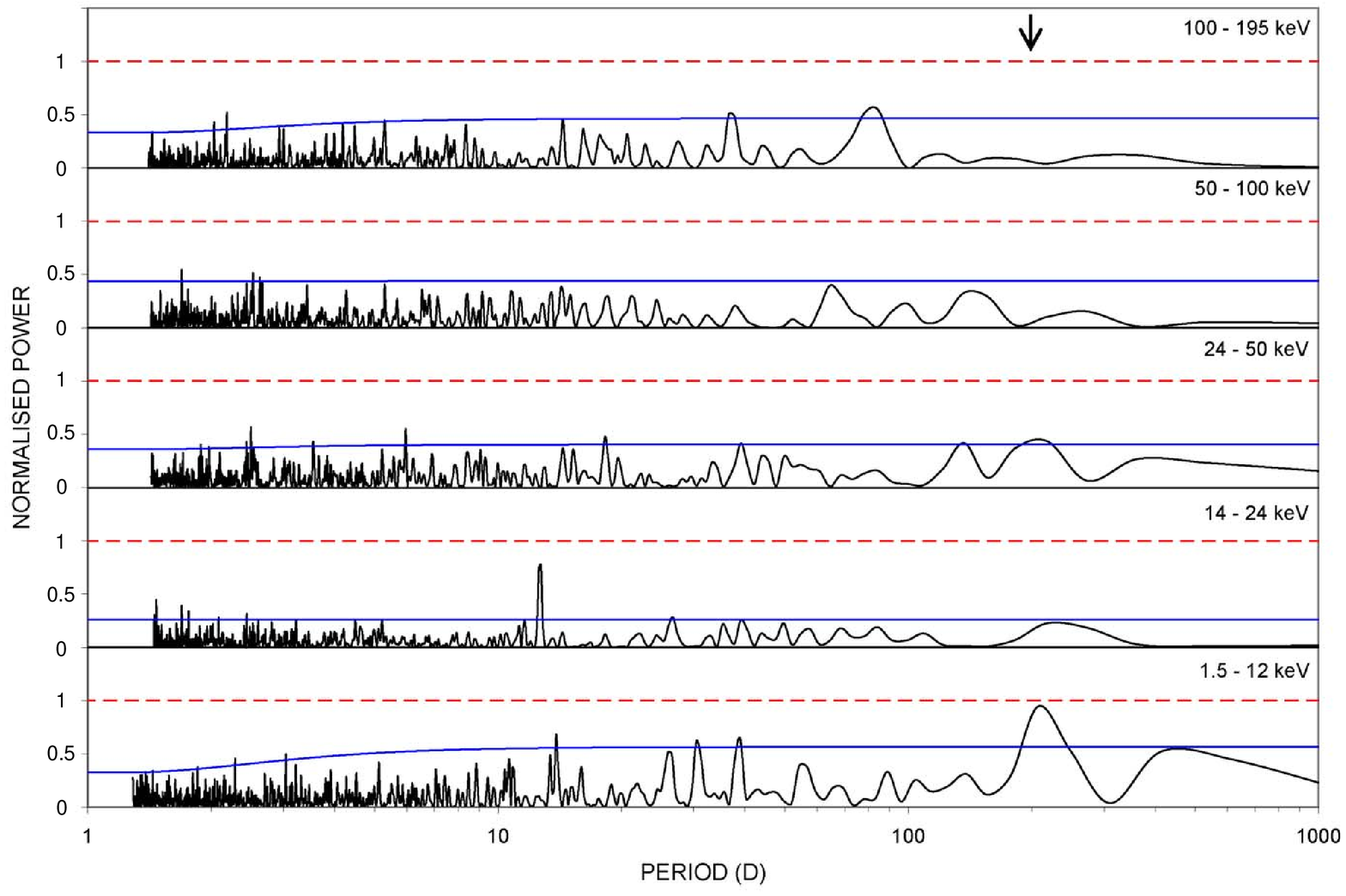}
\caption{4U 1916-053: normalised power spectra of the 1.5 -- 12
keV \emph{RXTE} ASM light curve (bottom panel) and the 14 -- 24
keV, 24 -- 50 keV, 50 -- 100 keV and 100 -- 195 keV \emph{Swift}
BAT light curves (top panels). The arrow indicates the location of
the previously reported super-orbital period of 199
d.}\label{4U19}
\end{figure*}

\begin{figure*}
\includegraphics[width=140mm]{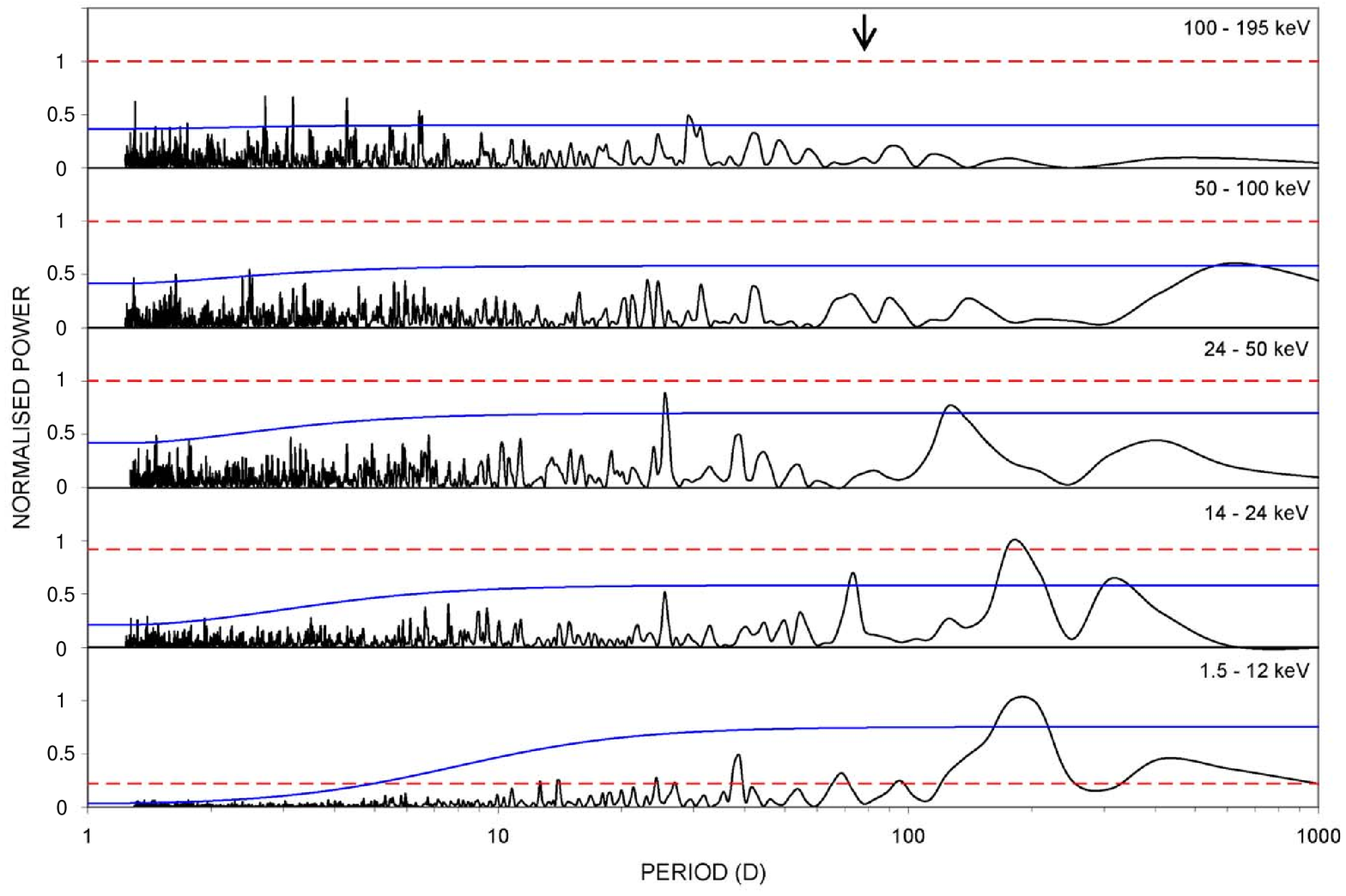}
\caption{Cyg X-2: normalised power spectra of the 1.5 -- 12 keV
\emph{RXTE} ASM light curve (bottom panel) and the 14 -- 24 keV,
24 -- 50 keV, 50 -- 100 keV and 100 -- 195 keV \emph{Swift} BAT
light curves (top panels). The arrow indicates the location of the
previously reported super-orbital period of 78 d.}\label{CygX2}
\end{figure*}

\begin{figure*}
\includegraphics[width=140mm]{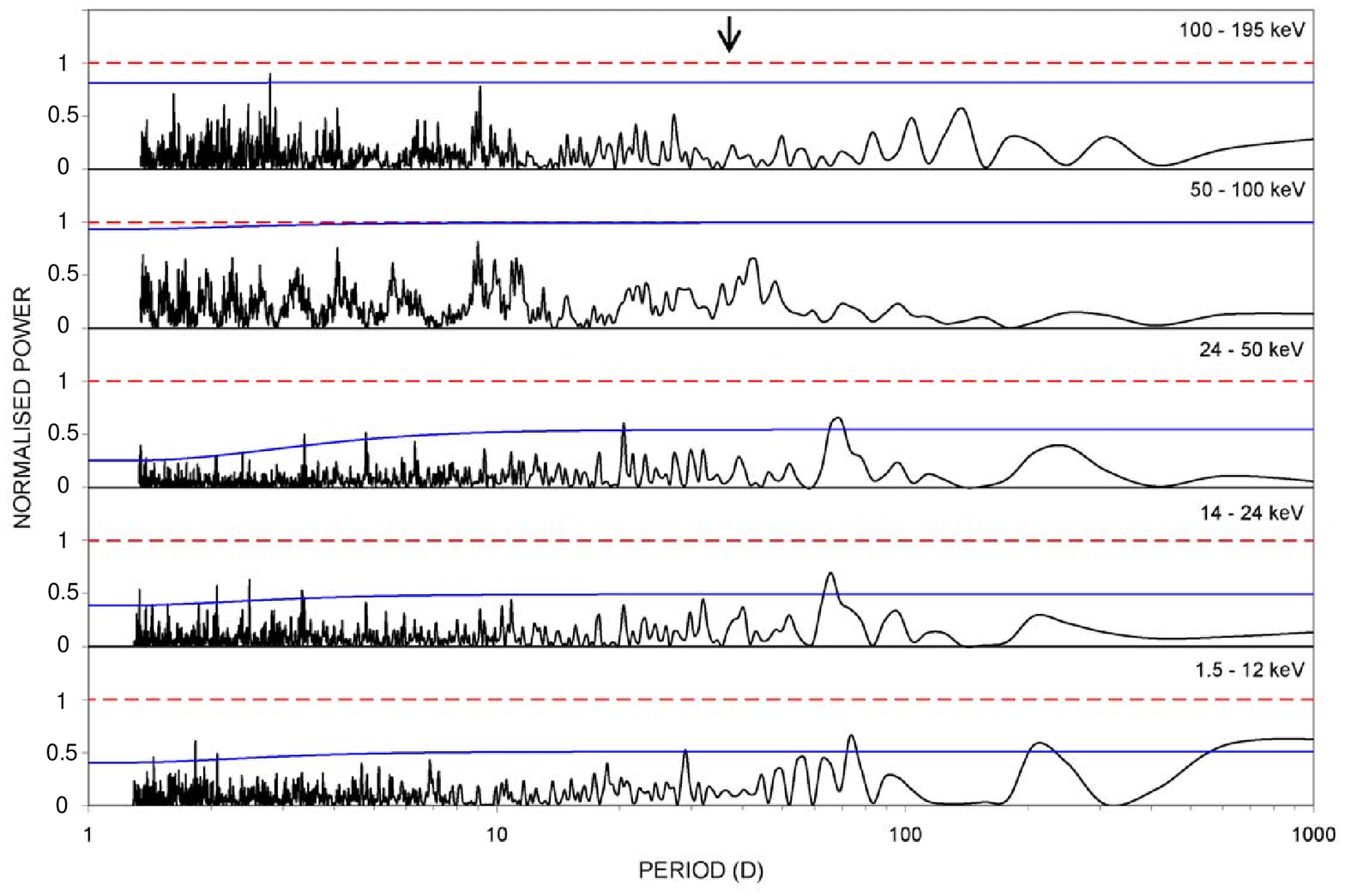}
\caption{Sco X-1: normalised power spectra of the 1.5 -- 12 keV
\emph{RXTE} ASM light curve (bottom panel) and the 14 -- 24 keV,
24 -- 50 keV, 50 -- 100 keV and 100 -- 195 keV \emph{Swift} BAT
light curves (top panels). The arrow indicates the location of the
previously reported super-orbital periods of 37 d.}\label{ScoX1}
\end{figure*}


\label{lastpage}


\begin{thebibliography}{99}

\bibitem[\protect\citeauthoryear{Andrew \& Purton}{1968}]{an68}
Andrew B. H., Purton C. R., 1968, Nat, 218, 855

\bibitem[\protect\citeauthoryear{Barret}{2001}]{bar01} Barret
D., 2001, AdSpR, 28, 307

\bibitem[\protect\citeauthoryear{Barthelmy et al.}{2005}]{bar05}
Barthelmy S.~D., et al., 2005, SSRv, 120, 143

\bibitem[\protect\citeauthoryear{Becker et al.}{1977}]{be77}
Becker R. H., Smith B. W., Swank J. H., Boldt E. A., Holt S. S.,
Pravdo S. H., Serlemitsos P. J., 1977, ApJ, 216, L101

\bibitem[\protect\citeauthoryear{Berendsen et
al.}{2000}]{ber00} Berendsen S.~G.~H., Fender R., Kuulkers E.,
Heise J., van der Klis M., 2000, MNRAS, 318, 599


\bibitem[\protect\citeauthoryear{Bloser et al.}{2000a}]{bl00a}
Bloser P. F., Grindlay J. E., Barret D., Boirin L., 2000a, ApJ,
542, 989

\bibitem[\protect\citeauthoryear{Bloser et al.}{2000b}]{bl00b}
Bloser P.~F., Grindlay J.~E., Kaaret P., Zhang W., Smale A.~P.,
Barret D., 2000b, ApJ, 542, 1000


\bibitem[\protect\citeauthoryear{Bradshaw, Fomalont \& Geldzahler}{Bradshaw et al.}{1999}]{br99}
Bradshaw C. F., Fomalont E. B., Geldzahler B. J., 1999, ApJ, 512,
L121

\bibitem[\protect\citeauthoryear{Byram, Chubb \& Friedman}{Byram et al.}{1966}]{by66}
Byram E. T., Chubb T. A., Friedman H., 1966, Sci, 152, 66

\bibitem[\protect\citeauthoryear{Casares, Charles \& Kuulkers}{Casares et al.}{1998}]{ca98}
Casares J., Charles P. A., Kuulkers E., 1998, ApJ, 493, L39

\bibitem[\protect\citeauthoryear{Casares et
al.}{2006}]{ca06} Casares J., Cornelisse R., Steeghs D., Charles
P.~A., Hynes R.~I., O'Brien K., Strohmayer T.~E., 2006, MNRAS,
373, 1235

\bibitem[\protect\citeauthoryear{Corbel et al.}{2003}]{cor03}
Corbel S., Nowak M.~A., Fender R.~P., Tzioumis A.~K., Markoff S.,
2003, A\&A, 400, 1007

\bibitem[\protect\citeauthoryear{Chou \& Grindlay}{2001}]{ch01} Chou Y., Grindlay J. E., 2001,
ApJ, 563, 934

\bibitem[\protect\citeauthoryear{Chou, Grindlay \& Bloser}{2001}]{ch01b} Chou Y., Grindlay J. E., Bloser P. F. 2001,
ApJ, 549, 1135

\bibitem[\protect\citeauthoryear{Clarkson et al.}{2003a}]{cl03a} Clarkson W.~I.,
Charles P.~A., Coe M.~J., Laycock S., Tout M.~D., Wilson C.~A.,
2003a, MNRAS, 339, 447

\bibitem[\protect\citeauthoryear{Clarkson et
al.}{2003b}]{cl03b} Clarkson W.~I., Charles P.~A., Coe M.~J.,
Laycock S., 2003b, MNRAS, 343, 1213

\bibitem[\protect\citeauthoryear{Corbet, Markwardt \& Tueller}{2007}]{co07}
Corbet R. H. D., Markwardt C. B., Tueller J., 2007, ApJ, 655, 458

\bibitem[\protect\citeauthoryear{D'Amico et al.}{2001}]{da01}
D'Amico F., Heindl W. A., Rothschild R. E., Gruber D. E., 2001,
ApJ, 547, L147

\bibitem[\protect\citeauthoryear{Di Salvo, Robba \& Stella}{Di Salvo et al.}{2002}]{di02} Di
Salvo T., Robba N., Stella L., 2002, MmSAI, 73, 1082

\bibitem[\protect\citeauthoryear{Di Salvo et al.}{2006}]{di06} Di Salvo T., Goldoni P., Stella L., van der Klis M., Bazzano A., Burderi L., Farinelli R., Frontera F., Israel G. L., M\'{e}ndez M., Mirabel I. F., Robba N. R., Sizun P., Ubertini P., Lewin W. H.
G., 2006, ApJ, 649, L91

\bibitem[\protect\citeauthoryear{Farrell, O'Neill \& Sood}{Farrell et al.}{2005}]{fa05} Farrell S. A., O'Neill P. M, Sood
R. K., 2005, PASA, 22, 267


\bibitem[\protect\citeauthoryear{Farrell et
al.}{2008}]{fa07} Farrell S.~A., Sood R.~K., O'Neill P.~M.,
Dieters S., 2008, MNRAS, 389, 608

\bibitem[\protect\citeauthoryear{Fender}{2006}]{fe06} Fender
R., 2006, in Lewin W., van der Klis M., eds., Compact Stellar
X-ray Sources. Cambridge University Press, Cambridge, p. 381

\bibitem[\protect\citeauthoryear{Fiocchi et al.}{2006}]{fi06}
Fiocchi M., Bazzano A., Ubertini P., Jean P., 2006, ApJ, 651, 416

\bibitem[\protect\citeauthoryear{Fomalont, Geldzahler \& Bradshaw}{Fomalont et al.}{2001}]{fo01}
Fomalont E. B., Geldzahler B. J., Bradshaw C. F., 2001, ApJ, 553,
L27

\bibitem[\protect\citeauthoryear{Galloway et
al.}{2006}]{gal06} Galloway D.~K., Psaltis D., Muno M.~P.,
Chakrabarty D., 2006, ApJ, 639, 1033


\bibitem[\protect\citeauthoryear{Giacconi et al.}{1962}]{gi62} Giacconi R., Gursky H., Paolini F., Rossi B., 1962, Phys. Rev.
Lett., 9, 439

\bibitem[\protect\citeauthoryear{Giacconi et al.}{1974}]{gi74} Giacconi R., Murray
S., Gursky H., Kellogg E., Schreier E., Matilsky T., Koch D.,
Tananbaum H., 1974, ApJS, 27, 37

\bibitem[\protect\citeauthoryear{Gottlieb, Wright \& Liller}{Gottlieb et
al.}{2001}]{go75} Gottlieb E. W., Wright E. L., Liller W., 1975,
ApJ, 195, L33

\bibitem[\protect\citeauthoryear{Grindlay et al.}{1976}]{gr76} Grindlay J. E., Gursky H., Schnopper H.,
Parsignault D. R., Heise J., Brinkman A. C., Schrijver J., 1976,
ApJ, 205, L127

\bibitem[\protect\citeauthoryear{Grindlay et al.}{1988}]{gr88} Grindlay, J. E., Bailyn C. D., Cohn H., Lungger P. M., Thorstensen
J. R., Wegner G. 1988, ApJ, 334, L25

\bibitem[\protect\citeauthoryear{Grindlay}{1992}]{gr92} Grindlay, J. E., 1992, in \emph{Proc. 28th Yamada Conf., Frontiers of X-Ray Astronomy}, eds. Y. Tanaka \harvardand{} K. Koyama,
 Universal Academy,
 Tokyo, 69

\bibitem[\protect\citeauthoryear{Harmon et al.}{2004}]{har04}
Harmon B.~A., et al., 2004, ApJS, 154, 585

\bibitem[\protect\citeauthoryear{Hasinger \& van der
 Klis}{1989}]{ha89} Hasinger G., van der Klis M., 1989, A\&A, 225,
 79

\bibitem[\protect\citeauthoryear{Hoffman, Lewin \& Doty}{Hoffman et
al.}{1977}]{ho77} Hoffman J. A., Lewin W. H. G., Doty J., 1977,
ApJ, 217, L23

\bibitem[\protect\citeauthoryear{Homer et al.}{2001}]{ho01}
Homer L., Charles P.~A., Hakala P., Muhli P., Shih I.-C., Smale
A.~P., Ramsay G., 2001, MNRAS, 322, 827

\bibitem[\protect\citeauthoryear{Jonker \&
Nelemans}{2004}]{jon04} Jonker P.~G., Nelemans G., 2004, MNRAS,
354, 355


\bibitem[\protect\citeauthoryear{Kong, Charles \& Kuulkers}{Kong et al.}{1998}]{ko98} Kong A. K. H., Charles P. A., Kuulkers E., 1998, NewA,
3, 301

\bibitem[\protect\citeauthoryear{Kuulkers et al.}{2003}]{ku03} Kuulkers E., den Hartog P. R., in't Zand J. J. M., Verbunt F. W. M., Harris W. E., Cocchi
M., 2003, A\&A, 399, 663

\bibitem[\protect\citeauthoryear{Larwood et
al.}{1996}]{la97} Larwood J.~D., Nelson R.~P., Papaloizou
J.~C.~B., Terquem C., 1996, MNRAS, 282, 597

\bibitem[\protect\citeauthoryear{Lavagetto et al.}{2006}]{la06}
Lavagetto G., di Salvo T., Falanga M., Iaria R., Robba N. R.,
Burderi L., Lewin W. H. G., M\'{e}ndez M., Stella L., van der Klis
M., 2006, A\&A, 445, 1089

\bibitem[\protect\citeauthoryear{Lewin et al.}{1977}]{le77} Lewin W. H. G.,
Hoffman J. A., Doty J., Clark G. W., Swank J. H., Becker R. H.,
Pravdo S. H., Serlemitsos P. J., 1977, Nat, 267, 28

\bibitem[\protect\citeauthoryear{Lomb}{1975}]{lo75} Lomb N. R., 1975, Ap$\&$SS, 39, 447

\bibitem[\protect\citeauthoryear{Malzac}{2007}]{mal07} Malzac
J., 2007, Ap\&SS, 311, 149


\bibitem[\protect\citeauthoryear{Margon}{1984}]{ma84} Margon
B., 1984, ARA\&A, 22, 507

\bibitem[\protect\citeauthoryear{Migliari \&
Fender}{2006}]{mi06} Migliari S., Fender R.~P., 2006, MNRAS, 366,
79

\bibitem[\protect\citeauthoryear{Migliari et
al.}{2007}]{mig07} Migliari S., et al., 2007, ApJ, 671, 706


\bibitem[\protect\citeauthoryear{Mirabel \& Rodrigues}{2003}]{mi03} Mirabel
I. F., Rodrigues I., 2003, A\&A, 398, L25

\bibitem[\protect\citeauthoryear{Ogilvie \& Dubus}{2001}]{og01} Ogilvie G.~I., Dubus G., 2001, MNRAS, 320, 485

\bibitem[\protect\citeauthoryear{Orosz \& Kuulkers}{1999}]{or99}
Orosz J. A., Kuulkers E., 1999, MNRAS, 305, 1320

\bibitem[\protect\citeauthoryear{Paul \& Kitamoto}{2002}]{pa02} Paul B.,
Kitamoto S., 2002, JApA, 23, 33

\bibitem[\protect\citeauthoryear{Paul, Kitamoto \& Makino}{Paul et al.}{2000}]{pa00} Paul B.,
Kitamoto S., Makino F., 2000, ApJ, 528, 410

\bibitem[\protect\citeauthoryear{Pedersen, van Paradijs \& Lewin}{Pedersen et al.}{1981}]{pe81}
Pedersen H., van Paradijs J., Lewin W. H. G., 1981, Nat, 294, 725

\bibitem[\protect\citeauthoryear{Peele \& White}{1996}]{pe96}
Peele A. G., White N. E., 1996, IAU Circ., 6524

\bibitem[\protect\citeauthoryear{Piraino, Santangelo \& Kaaret}{Piraino et al.}{2002}]{pi02}
Piraino S., Santangelo A., Kaaret P., 2002, ApJ, 567, 1091

\bibitem[\protect\citeauthoryear{Press \& Rybicki}{1989}]{pr89} Press W. H., Rybicki G. B., 1989, ApJ, 338, 277

\bibitem[\protect\citeauthoryear{Priedhorsky \& Terrell}{1984a}]{pr84a} Priedhorsky W., Terrell
J., 1984a, ApJ, 280, 661

\bibitem[\protect\citeauthoryear{Priedhorsky \& Terrell}{1984b}]{pr84b} Priedhorsky W., Terrell
J., 1984b, ApJ, 284, L17

\bibitem[\protect\citeauthoryear{Psaltis}{2006}]{ps06} Psaltis D., 2006, Compact stellar X-ray sources, 1

\bibitem[\protect\citeauthoryear{Rappaport et al.}{1987}]{ra87} Rappaport S., Ma C. P., Joss P. C., Nelson L.
A., 1987, ApJ, 322, 842

\bibitem[\protect\citeauthoryear{Rutten, van Paradijs \& Tinbergen}{Rutten et al.}{1992}]{ru92} Rutten R. G. M., van Paradijs J. Tinbergen J., 1992, A$\&$A, 260, 213

\bibitem[\protect\citeauthoryear{Scargle}{1982}]{sc82} Scargle J. D.,
1982, ApJ, 263, 835

\bibitem[\protect\citeauthoryear{Scargle}{1989}]{sc89} Scargle J. D., 1989, ApJ, 343, 874

\bibitem[\protect\citeauthoryear{Schulz \& Mudelsee}{2002}]{sc02} Schulz
M., Mudelsee M., 2002, CG, 28, 421

\bibitem[\protect\citeauthoryear{Shih et al.}{2005}]{sh05} Shih I.
C., Bird A. J., Charles P. A., Cornelisse R., Tiramani D., 2005,
MNRAS, 361, 602

\bibitem[\protect\citeauthoryear{\v{S}imon}{2005}]{si05} \v{S}imon V.,
2005, A\&A, 436, 263

\bibitem[\protect\citeauthoryear{Smale}{1998}]{sm98} Smale A. P.,
1998, ApJ, 498, L141

\bibitem[\protect\citeauthoryear{Smale \& Lochner}{1992}]{sm92} Smale A. P., Lochner J. C., 1992, ApJ, 395, 582

\bibitem[\protect\citeauthoryear{Sood et al.}{2006}]{so06} Sood
R. K., Farrell S. A., O'Neill P. M, Manchanda R. K., Ashok N. M.,
2006, ASR, 38, 2779

\bibitem[\protect\citeauthoryear{Sood et al.}{2007}]{so07} Sood
R. K., Farrell S. A., O'Neill P. M, Dieters S., 2007, ASR, 40,
1528

\bibitem[\protect\citeauthoryear{Stairs, Lyne \& Shemar}{Stairs et al.}{2000}]{sta00} Stairs I. H., Lyne A. G., Shemar S. L., 2000, Nat, 406, 484

\bibitem[\protect\citeauthoryear{Stella, Priedhorsky \& White}{Stella et al.}{1987}]{st87} Stella L., Priedhorsky W., White N.
E., 1987, ApJ, 312, L17

\bibitem[\protect\citeauthoryear{Swank, Taam \& White}{Swank et al.}{1984}]{sw84}
Swank J. H., Taam R. E., White N. E., 1984, ApJ, 277, 274

\bibitem[\protect\citeauthoryear{Strickman \&
Barret}{2000}]{st00} Strickman M., Barret D., 2000, AIPC, 510, 222


\bibitem[\protect\citeauthoryear{Tarana et al.}{2007}]{ta07}
Tarana A., Bazzano A., Ubertini P., Zdziarski A. A., 2007, ApJ,
654, 494

\bibitem[\protect\citeauthoryear{Tr\"{u}mper et al.}{1986}]{tr86} Tr\"{u}mper J., Kahabka P., \"{O}gelman H., Pietsch W., Voges W., 1986,
ApJ, 300, L63

\bibitem[\protect\citeauthoryear{Vanderlinde, Levine \& Rappaport}{Vanderlinde et al.}{2003}]{va03}
Vanderlinde K. W., Levine A. M., Rappaport S. A., 2003, PASP, 115,
739

\bibitem[\protect\citeauthoryear{Weiler et al.}{1992}]{we92} Weiler K. W., van Dyk S. D., Pringle J. E., Panagia N., 1992,
ApJ, 399, 672

\bibitem[\protect\citeauthoryear{Wen et al.}{2006}]{we06}
Wen L., Levine A.~M., Corbet R.~H.~D., Bradt H.~V., 2006, ApJS,
163, 372

\bibitem[\protect\citeauthoryear{White \& Swank}{1982}]{wh82} White
N. E., Swank J. H., 1982, ApJ, 253, L61

\bibitem[\protect\citeauthoryear{Wijers \& Pringle}{1999}]{wij99}
Wijers R. A. M., Pringle J. E., 1999, MNRAS, 308, 207

\bibitem[\protect\citeauthoryear{Wijnands, Kuulkers \& Smale}{Wijnands et al.}{1996}]{wi96}
Wijnands R. A. D., Kuulkers E., Smale A. P., 1996, ApJ, 473, L45

\bibitem[\protect\citeauthoryear{Zdziarski,
Wen \& Gierli{\'n}ski}{Zdziarski et al.}{2007a}]{zd07a} Zdziarski
A.~A., Wen L., Gierli{\'n}ski M., 2007a, MNRAS, 377, 1006

\bibitem[\protect\citeauthoryear{Zdziarski et al.}{2007b}]{zd07b} Zdziarski A.~A., Gierli{\'n}ski
M., Wen L., Kostrzewa Z., 2007b, MNRAS, 377, 1017

\bibitem[\protect\citeauthoryear{Zhang et al.}{1998}]{zh98} Zhang W., Smale A. P., Strohmayer T. E., Swank J.
 H., 1998, ApJ, 500, L171

\end{thebibliography}
\end{document}